\documentclass[aps, prd, floats, floatfix,superscriptaddress, nofootinbib, twocolumn,showpacs]{revtex4-1}

\usepackage{graphicx}
\usepackage[usenames,dvipsnames]{color}
\usepackage[colorlinks, pdfborder={0 0 0}, plainpages=false]{hyperref}
\usepackage{amsmath, amssymb, amsfonts}
\usepackage{xspace} 
\usepackage{dcolumn}
\usepackage{bm}
\usepackage{multirow} 

\definecolor{CiteColor}{rgb}{0, 0.5, 0}
\hypersetup{citecolor=CiteColor} %
\definecolor{RefColor}{rgb}{0.55, 0, 0}
\hypersetup{linkcolor=RefColor} %

\newcommand{\T}{\rule{0pt}{3ex}}
\newcommand{\B}{\rule[-2ex]{0pt}{0pt}}

\usepackage{ulem}
\usepackage{amsmath}
\newcommand{\MIT}{\affiliation{Department of Physics and MIT Kavli Institute, 77 
Massachusetts Avenue, Cambridge, MA 02139}}
\newcommand{\Maryland}{\affiliation{Maryland Center for Fundamental
    Physics \& Joint Space-Science Institute, \\ Department of Physics, University of Maryland, College
    Park, MD 20742}}
\newcommand{\Radcliffe}{\affiliation{Radcliffe Institute for Advanced Study, Harvard University, 
8 Garden St., Cambridge, MA 02138}}
\newcommand{\Darth}{\affiliation{Department of Physics, University of Massachusetts Dartmouth, 
North Dartmouth, MA 02747}}

\definecolor {darkgreen}{rgb}{0.2,0.7,0.2}

\begin{document}

\title{Modeling multipolar gravitational-wave emission from small mass-ratio mergers}

\author{Enrico Barausse} \Maryland 
\author{Alessandra Buonanno} \Maryland \Radcliffe
\author{Scott A. Hughes} \MIT
\author{Gaurav Khanna} \Darth
\author{Stephen O'Sullivan} \MIT
\author{Yi Pan} \Maryland 

\begin{abstract}
Using the effective-one-body (EOB) formalism and a time-domain Teukolsky code, we generate inspiral, merger,
and ringdown waveforms in the small mass-ratio limit.  We use EOB inspiral and plunge trajectories to build
the Teukolsky equation source term, and compute full coalescence waveforms for a range of black hole spins.
By comparing EOB waveforms that were recently developed for comparable mass binary black holes to these
Teukolsky waveforms, we improve the EOB model for the $(2,2)$, $(2,1)$, $(3,3)$, and $(4,4)$ modes.  Our
results can be used to quickly and accurately extract useful information about merger waves for binaries
with spin, and should be useful for improving analytic models of such binaries. 
\end{abstract}

\date{\today \hspace{0.2truecm}}


\maketitle

\section{Introduction}
\label{sec:intro}

Since the numerical relativity breakthrough of 2005~\cite{Pretorius2005a,Campanelli2006a,Baker2006d}, 
there have been tremendous advances both in the computation of gravitational radiation from binary
black-hole systems, and in analytical modeling of this radiation using approximate techniques.
Despite rapid and ongoing advances, it remains a challenge for numerical relativity to quickly and
accurately compute models that span large regions of parameter space.  Extreme conditions such
as large spins and small mass ratios are particularly challenging, although there has been excellent
recent progress on these issues~\cite{DainEtAl:2008,Lovelace2010,LoustoZlowchower:2011}.

The effective-one-body (EOB) formalism~\cite{Buonanno99,Buonanno00, DJS00, Damour01c, Buonanno06,
  Damour:024009,DIN,Barausse:2009xi,Pan:2010hz,Nagar:2011fx,Barausse:2011ys}
makes it possible to analytically model the three main phases of binary black-hole evolution: inspiral,
plunge-merger, ringdown. EOB has been used to model the dynamics and gravitational-wave emission
from comparable-mass binaries~\cite{Buonanno-Cook-Pretorius:2007, Buonanno2007, Pan2007, Boyle2008a,
Buonanno:2009qa, Pan:2009wj, Damour2007a, DN2007b,DN2008,Damour2009a,Pan:2011gk},
extreme mass-ratio inspiraling binaries~\cite{Yunes:2009ef,AmaroSeoane:2010ub,Yunes:2010zj} (neglecting
conservative self-force effects), and small mass-ratio non-spinning binaries~\cite{Nagar:2006xv,
Damour2007,Bernuzzi:2010ty,Bernuzzi:2010xj,Bernuzzi:2011aj}.  In order to study the transition from
inspiral to plunge-merger and ringdown, Refs.~\cite{Nagar:2006xv,Damour2007} suggested combining EOB
with black hole perturbation theory. Concretely, they used EOB in order to compute the trajectory
followed by an object spiraling and plunging into a much larger black hole, and then used that trajectory
to describe the source for the time-domain Regge-Wheeler-Zerilli (RWZ) equation {\cite{ReggeWheeler1957,
Zerilli1970}} describing metric perturbations to Schwarzschild black holes.  They were then able to compute
the full RWZ coalescence waveform and to compare with the EOB model.  This was used in
Ref.~\cite{Bernuzzi:2010ty} to produce gravitational modes beyond the leading $(2,2)$ mode and compute
the recoil velocity. More recently, Refs.~\cite{Bernuzzi:2010xj,Bernuzzi:2011aj} have used information
from the RWZ modes to improve the modeling of the subleading EOB modes.  A particularly beautiful
feature of Ref.\ {\cite{Bernuzzi:2011aj}} is the use, for the first time, of hyperboloidal slicings in
such an analysis.  This effectively compactifies the computational domain so that waveforms at future null
infinity can be read out of the numerical calculation with great accuracy.

References {\cite{skh07,skh08,Sundararajan:2010sr}} developed a time-domain computational
framework based on the Teukolsky equation {\cite{Teukolsky1972}}, which describes curvature perturbations
of rotating (Kerr) black holes.  The goal of these papers has been to understand gravitational waves
produced by physically reasonable but otherwise arbitrary trajectories of small bodies bound to rotating
black holes (such as slowly inspiraling orbits, or trajectories that plunge into the hole's event
horizon).  This has been used to understand the small-mass ratio limit of merging black holes,
studying for example the dependence of recoil velocity on black hole spin in this limit. This Teukolsky code
has been optimized to make effective use of modern many-core processor architectures, such as Graphics
Processing Units (GPUs) {\cite{McKenKha2010}}.

In this paper, we combine EOB with the time-domain Teukolsky code developed in
Refs.~\cite{skh07,skh08,Sundararajan:2010sr} to extend the ideas of Refs.~\cite{Nagar:2006xv,Damour2007,
Bernuzzi:2010xj,Bernuzzi:2011aj} in several directions.  Our primary extension is, for the first time, producing
full coalescence waveforms describing inspiral, merger, and ringdown for quasi-circular equatorial orbits
in the Kerr spacetime.  The energy flux we use in the EOB equations of motion comes from the factorized
resummed waveforms of Refs.\ {\cite{DIN,Pan:2010hz}}.  For the Schwarzschild limit, we model
analytically three subleading modes [$(2,1)$, $(3,3)$, and $(4,4)$] plus the dominant $(2,2)$ mode,
finding useful information about the plunge-merger, which we use to improve the comparable mass EOB model
described in Ref.\ {\cite{Pan:2011gk}}.  For more general spins, we calibrate the leading EOB mode
for spins $a/M = -0.9$, $-0.5$, $0.5$, and $0.7$.  We also extract some information regarding subleading
modes, and regarding the high prograde spin $a/M = 0.9$.  These results for spinning binaries provide
valuable input for improving the spinning EOB model of Refs.~\cite{Barausse:2009aa,Barausse:2009xi}, as
well as the spinning EOB waveforms of Refs.~\cite{Pan:2009wj,Taracchini:2011}.  This will in turn make it
possible to develop models that can cover a much larger region of parameter space, including higher modes
and extreme spins. 

Several other groups have also been using perturbation theory tools recently to improve our understanding
for comparable mass and intermediate mass ratio binaries.  For example, in Refs.~\cite{Lousto:2010ut,
LoustoZlowchower:2011,Nakano:2011pb} the authors directly employ a moving-puncture trajectory (or 
a post-Newtonian--inspired fit to it) in the RWZ equation. They compare the resulting RWZ waveform with the 
results of full numerical relativity calculation for
mass ratios $1/15$ and $1/10$, finding good agreement.  Another recent suggestion is the hybrid approach
of Ref.~\cite{Han:2011qz}, in which inspiral-plunge intermediate-mass black-hole waveforms are computed
by evolving the EOB equations of motion augmented by the perturbation-theory energy flux.  An important
issue in all attempts to model binary coalescence with perturbation theory is the computation of the
so-called excitation coefficients, or more generally the question of which fundamental frequencies
contribute to the radiation.  In this context, perturbation-theory calculations are offering new
insights~\cite{MinoBrink2008,Zimmerman:2011dx,Hadar:2009ip,Hadar:2011vj}.   

The remainder of this paper is organized as follows.  We begin in Sec.~\ref{sec:analytics} by reviewing
the EOB formalism for a test particle moving along quasi-circular, equatorial orbits around a Kerr black
hole.  We then describe (Sec.~\ref{sec:num}) the time-domain Teukolsky equation calculation we use to
compute the gravitational radiation emitted from a test particle that follows our EOB-generated trajectory.
This section discusses in some detail numerical errors which arise from finite-difference discretization,
and from the extrapolation procedure by which we estimate our waves at future null infinity.  Since we
began this analysis, the Teukolsky code we use has been upgraded to use the hyperboloidal layer method
{\cite{ZenKha2011}}.  This upgrade came too late to be used throughout our analysis, but has been used to
spot check our estimates of this extrapolation error.

Our results are presented in Sec.\ {\ref{sec:results}}.  We begin by comparing the leading and three
subleading Teukolsky modes with $a/M = 0$ to the corresponding EOB modes calibrated to non-spinning
comparable mass binaries {\cite{Pan:2011gk}}.  We then improve the non-spinning EOB model by including
some features we find in our test-particle-limit calculation.  We next calibrate the leading $(2,2)$
EOB waveform with our Teukolsky-equation results for spins $a/M = -0.9$, $-0.5$, $0.5$, and $0.7$.  We
conclude our results by discussing the challenges of calibrating subleading modes and of modeling extreme
spin configurations, such as ones with $a/M \geq 0.9$.  Section {\ref{sec:concl}} summarizes our
main conclusions, and outlines some plans for future work.  A particularly important goal for the future
will be to move beyond equatorial configurations, modeling the important case of binaries with misaligned
spins and orbits.

Throughout this paper, we use geometric units with $G=c=1$.

\section{Dynamics and waveforms using the effective-one-body formalism}
\label{sec:analytics}

The Hamiltonian of a non-spinning test-particle of mass $\mu$ orbiting a Kerr black hole of mass $M$ and 
intrinsic angular momentum (or spin) per unit mass $a$ is
\begin{equation}\label{hatHreal}
 H = \beta^i \, p_i + \alpha\,\sqrt{\mu^2 + \gamma^{ij}\,p_i\,p_j}\,,
\end{equation} 
where the indices $i,j$ label spatial directions, and the functions introduced here are given by
\begin{subequations}
\begin{eqnarray}
\label{alpha}
\alpha &=& \frac{1}{\sqrt{-g^{tt}}}\,,\\
\beta^i &=& \frac{g^{ti}}{g^{tt}}\,,\\
\gamma^{ij} &=& g^{ij}-\frac{g^{ti}\,g^{tj}}{g^{tt}}\,;
\label{gamma}
\end{eqnarray}
\end{subequations}
$t$ is the time index, and $g_{\mu\nu}$ is the Kerr metric.  Working in Boyer-Lindquist coordinates
$(t,r,\theta,\phi)$ and restricting ourselves to the equatorial plane $\theta = \pi/2$, the
relevant metric components read
\begin{subequations}
\begin{eqnarray}
\label{def_metric_in}
g^{tt} &=& -\frac{\Lambda}{r^2\,\Delta}\,,\\
g^{rr} &=& \frac{\Delta}{r^2}\,,\\
g^{\phi\phi} &=& \frac{1}{\Lambda}
\left(-\frac{4a^2\,M^2}{\Delta} + r^2\right)\,,\label{eq:gff}\\
g^{t\phi}&=&-\frac{2 a \,M}{r\,\Delta}\,,\label{def_metric_fin}
\end{eqnarray}
\end{subequations}
where we have introduced the metric potentials
\begin{eqnarray}
\label{Delta}
\Delta &=& r^2 - 2Mr - a^2\,, \\
\Lambda &=& (r^2 + a^2)^2 - a^2\,\Delta\,.
\label{Lambda}
\end{eqnarray}
We replace the radial momentum $p_r$ with $p_{r^*}$, the momentum conjugate to the {\it tortoise}
radial coordinate $r^*$.  The tortoise coordinate is related to the Boyer-Lindquist $r$ by
\begin{equation}\label{tortoise}
dr^*=\frac{r^2+a^2}{\Delta}\,dr \,.
\end{equation}
Since $p_r$ diverges at the horizon while $p_{r^*}$ does not, this replacement improves the numerical
stability of the Hamilton equations
\begin{subequations} \label{eq-eob}
\begin{align}
\frac{dr}{d {t}} &= \frac{\Delta}{r^2+a^2}\frac{\partial {H}}{\partial p_{r^*}}(r,p_{r^*},p_\phi)\,, \label{eq:eobhamone} \\
\frac{d \phi}{d {t}}  &= M\,\Omega =\frac{\partial {H}}{\partial p_\phi}(r,p_{r^*},p_\phi)\,, 
   \label{eq:eobhamtwo}\\
\frac{d p_{r^*}}{d {t}} &=-\frac{\Delta}{r^2+a^2}\frac{\partial {H}}
  {\partial r}(r,p_{r^*},p_\phi) +{}^{\rm nK}{\cal F}_\phi\,\frac{p_{r^*}}{p_\phi}\,, \label{eq:eobhamthree}\\
  \frac{d p_\phi}{d {t}} &={}^{\rm nK}{\cal F}_\phi\,.
  \label{eq:eobhamfour}
\end{align}
\end{subequations}
Our trajectory is produced by integrating these equations using initial conditions that specify a
circular orbit.  We typically find in our evolutions a small residual eccentricity on the order of
$3\times 10^{-4}$.

In Eqs.\ \eqref{eq:eobhamone}--\eqref{eq:eobhamfour}, radiation-reaction effects are included following
the EOB formalism.  For the $\phi$ component of the radiation-reaction force we use the non-Keplerian (nK)
force
\begin{equation}
{}^{\rm nK}{\cal F}_\phi = -\frac{1}{\nu\,v_\Omega^3}\,\frac{dE}{dt}\,,
\end{equation}
where $v_\Omega \equiv (M\,\Omega)^{1/3}$, and $dE/dt$ is the energy flux for quasi-circular
orbits obtained by summing over gravitational-wave modes $(l,m)$. We use  
\begin{equation}\label{resflux}
\frac{dE}{dt}=\frac{1}{16\pi}\,\sum_{\ell=2}^8\,\sum_{m=-\ell}^{\ell}m^2\,v_\Omega^6\,\left|h_{\ell m}\right|^2\,.
\end{equation}
The non-Keplerian behavior of the radiation-reaction force is implicitly introduced through the
definition of $h_{\ell m}$.  To describe the inspiral and plunge dynamics, we use the modes
\begin{equation}\label{hip}
h^{\rm insp-plunge}_{\ell m} = h^{\rm F}_{\ell m}\,N_{\ell m}\,.
\end{equation}
The coefficients $N_{\ell m}$ describe effects that go beyond the quasi-circular assumption and 
will be defined below [see Eq.~(\ref{Nlm})].  The factors $h^{\rm F}_{\ell m}$
are the factorized resummed modes, and are given by~\cite{DIN} 
\begin{equation}\label{hlm}
h^{\rm F}_{\ell m}=h_{\ell m}^{(N,\epsilon)}\,\hat{S}^{(\epsilon)}\,T_{\ell m}\,
e^{i\delta_{\ell m}}\,\left(\rho_{\ell m}\right)^\ell\,.
\end{equation}
Here, $\epsilon = \pi(\ell + m)$ is the parity of the multipolar waveform.  The leading term
in Eq.~(\ref{hlm}), $h_{\ell m}^{(N,\epsilon)}$, is the Newtonian contribution
\begin{equation}\label{hlmNewt}
h_{\ell m}^{(N,\epsilon)}=\frac{M\nu}{\cal{R}}\,n_{\ell m}^{(\epsilon)}\,c_{\ell+\epsilon}(\nu)\,V^{\ell}_\phi\,
Y^{\ell-\epsilon,-m}\,\left(\frac{\pi}{2},\phi\right)\,,
\end{equation}
where $\cal{R}$ is distance from the source, $Y^{\ell m}(\theta,\phi)$ are the scalar spherical
harmonics, and the functions $n_{\ell m}^{(\epsilon)}$ and $c_{\ell+\epsilon}(\nu)$ are given in Eqs.~(4a),
(4b) and (5) of Ref.~\cite{Pan2010hz} with $\nu=\mu/M$.  For reasons that we will explain in
Sec.~\ref{sec:results_original}, we choose
\begin{subequations}
\label{Veq}
\begin{eqnarray}
\label{unchanged}
V^{\ell}_\phi &=& v_\phi^{(\ell+\epsilon)} \qquad \qquad \; (\ell,m) \neq (2,1)\,, (4,4)\,,\\
\label{change}
V^{\ell}_\phi &=& \frac{1}{r_\Omega}\,v_\phi^{(\ell+\epsilon-2)} \qquad (\ell,m) = (2,1)\,, (4,4)\,.
\end{eqnarray}
\end{subequations}
The quantities $v_\phi$ and $r_\Omega$ introduced here are defined by
\begin{equation}
\label{vPhi}
v_\phi\equiv M\Omega\,r_\Omega \equiv M\Omega\left[\left(r/M\right)^{3/2}+a/M\right]^{2/3}\,.
\end{equation}
The function $\hat{S}^{(\epsilon)}$ in Eq.~(\ref{hlm}) is given by
\begin{equation}
\hat{S}^{(\epsilon)}(r,p_{r^*},p_\phi) = 
\begin{cases}
H(r,p_{r^*},p_\phi)\,, & \epsilon = 0\,, \\
L=p_\phi\,v_\Omega\,, & \epsilon = 1\,. 
\end{cases}
\end{equation}
The factor $T_{\ell m}$ in Eq.~(\ref{hlm}) resums the leading order logarithms 
of tail effects, and is given by
\begin{eqnarray}
T_{\ell m} &=&\frac{\Gamma(\ell+1-2i\,m\,M\,\Omega)}{\Gamma(\ell+1)}\times \nonumber \\
&& e^{\pi\,m\,M\,\Omega}\,e^{2i\,m\,M\,\Omega\,\log(2\,m\,\Omega\,r_0)}\,,
\end{eqnarray}
where $r_0=2M/\sqrt{e}$~\cite{Pan2010hz} and $\Gamma(z)\equiv\int_0^\infty t^{z-1}e^{-t}\,dt$ is the complex gamma function. The factor $e^{i\delta_{\ell m}}$ in Eq.~(\ref{hlm}) is a
phase correction due to subleading order logarithms; $\delta_{\ell m}$ is computed using
Eqs.~(27a)--(27i) of Ref.\ {\cite{Pan2010hz}}.  The factor $(\rho_{\ell m})^{\ell}$ in
Eq.~(\ref{hlm}) collects the remaining post-Newtonian terms, and is computed using
Eqs.~(29a)--(29i) and Eqs.~(D1a)--(D1m) of Ref.~\cite{Pan2010hz}.

Finally, the function $N_{\ell m}$ entering Eq.~(\ref{hip}) is given by
\begin{eqnarray}\label{Nlm}
N_{\ell m} &=& \left [ 1 + a^{h_{\ell m}}_{1}\,
\frac{p_{r^*}^{2}}{(r \,\Omega)^{2}} + a^{h_{\ell m}}_{2}\,\frac{p_{r^*}^{2}}{(r\,\Omega)^{2}}\frac{M}{r} \right. 
\nonumber \\
&& \left. + a^{h_{\ell m}}_{3}\,\frac{p_{r^*}^{2}}{(r\,\Omega)^{2}}\left (\frac{M}{r}\right)^{3/2} 
+ a^{h_{\ell m}}_{4}\,\frac{p_{r^*}^{2}}{(r\,\Omega)^{2}}\left (\frac{M}{r}\right)^2 \right. 
\nonumber \\
&& \left. + a^{h_{\ell m}}_{5}\,\frac{p_{r^*}^{2}}{(r\,\Omega)^{2}}\left (\frac{M}{r}\right)^{5/2}
\right ]\times 
\nonumber \\ 
&&\exp\left[i\,\left(b^{h_{\ell m}}_{1}\,\frac{p_{r^*}}{r \,\Omega}+b^{h_{\ell m}}_{2}\,\frac{p_{r^*}^3}{r \,\Omega}\right.\right.\nonumber\\
&&\left.\left.\quad\quad+b^{h_{\ell m}}_{3}\,\sqrt{\frac{M}{r}}\frac{p_{r^*}^3}{r \,\Omega}+b^{h_{\ell m}}_{4}\,\frac{M}{r}\frac{p_{r^*}^3}{r \,\Omega}\right)\right]\,
\end{eqnarray}
where the quantities $a^{h_{\ell m}}_{i}$ and $b^{h_{\ell m}}_{i}$ are non-quasicircular (NQC) orbit
coefficients. We will explain in detail how these coefficients are fixed in Sec.~\ref{sec:results}.

We conclude this section by describing how we build the final merger-ringdown portion of the EOB
waveform.  For each mode $(\ell,m)$ we have
\begin{equation}
\label{RD}
h_{\ell m}^{\rm merger-RD}(t) = \sum_{n=0}^{N-1} A_{\ell mn}\,
e^{-i\sigma_{\ell mn} (t-t_{\rm match}^{\ell m})},
\end{equation}
where $n$ labels the overtone of the Kerr quasinormal mode (QNM), $N$ is the number of overtones
included in our model, and $A_{\ell mn}$ are complex amplitudes to be determined by a matching procedure
described below. The complex frequencies $\sigma_{\ell mn} = \omega_{\ell m n} - i/\tau_{\ell m n}$, where the
quantities $\omega_{\ell m n}>0$ are the oscillation frequencies and $\tau_{\ell m n}>0$ are the decay times, 
are known functions of the final black-hole mass
and spin and can be found in Ref.~\cite{Berti2006a}. In this paper, we model the ringdown modes as a 
linear combination of eight QNMs (i.e., $N =8$).

The complex amplitudes $A_{\ell mn}$ in Eq.~(\ref{RD}) are determined by matching the merger-ringdown
waveform (\ref{RD}) with the inspiral-plunge waveform (\ref{hip}). In order to do this, $N$ independent
complex equations needs to be specified throughout the comb of width $\Delta t_\mathrm{match}^{\ell m}$.
Details on the procedure are given in Ref.~\cite{Pan:2011gk}.  The full inspiral(-plunge)-merger-ringdown
waveform is then given by
\begin{equation}
\label{eobfullwave}
h_{\ell m} = h_{\ell m}^{\rm insp-plunge}\,
\theta(t_{\rm match}^{\ell m} - t) + 
h_{\ell m}^{\rm merger-RD}\,\theta(t-t_{\rm match}^{\ell m})\,.
\end{equation}

In this analysis, we focus on waveforms emitted by a test-particle of mass $\mu$ orbiting a
Kerr black hole.  Thus, we shall set to zero terms proportional to $\nu = \mu/M$ in Eq.~(\ref{hlm}), 
excepting the leading $\nu$ term in Eq.\ (\ref{hlmNewt}).  Throughout this paper we restrict ourselves
to the case $\nu = 10^{-3}$.

\section{The time-domain Teukolsky code}
\label{sec:num}

\subsection{Overview: The Teukolsky equation and its solution}
\label{sec:teukoverview}

The evolution of scalar, vector, and tensor perturbations of a Kerr black hole is described by the
Teukolsky master equation~\cite{Teukolsky1972}, which takes the following form in
Boyer-Lindquist coordinates:
\begin{eqnarray}
\label{teuk0}
&&
-\left[\frac{(r^2 + a^2)^2 }{\Delta}-a^2\sin^2\theta\right]
         \partial_{tt}\Psi
-\frac{4 M a r}{\Delta}
         \partial_{t\phi}\Psi \nonumber \\
&&- 2s\left[r-\frac{M(r^2-a^2)}{\Delta}+ia\cos\theta\right]
         \partial_t\Psi\nonumber\\  
&&
+\,\Delta^{-s}\partial_r\left(\Delta^{s+1}\partial_r\Psi\right)
+\frac{1}{\sin\theta}\partial_\theta
\left(\sin\theta\partial_\theta\Psi\right)+\nonumber\\
&& \left[\frac{1}{\sin^2\theta}-\frac{a^2}{\Delta}\right] 
\partial_{\phi\phi}\Psi +\, 2s \left[\frac{a (r-M)}{\Delta} 
+ \frac{i \cos\theta}{\sin^2\theta}\right] \partial_\phi\Psi  \nonumber\\
&&- \left(s^2 \cot^2\theta - s \right) \Psi = -4\pi (r^2 + a^2 \cos^2 \theta)\, T\;.
\end{eqnarray}
The coordinates, the mass $M$, the spin parameter $a$, and the
function $\Delta$ are as defined in the previous section.  The number
$s$ is the ``spin weight'' of the field.  When $s = \pm 2$, this
equation describes radiative degrees of freedom for gravity.  We focus
on the case $s = -2$, for which $\Psi = (r-ia\cos\theta)^4 \psi_4$,
where $\psi_4$ is the Weyl curvature scalar that characterizes
outgoing gravitational waves.

To solve Eq.\ (\ref{teuk0}), we use an approach introduced by Krivan
et al.\ {\cite{Krivan1997}}. First, we change from radial coordinate
$r$ to tortoise coordinate $r^*$ [Eq.\ \eqref{tortoise}], and from
axial coordinate $\phi$ to $\tilde{\phi}$, defined by
\begin{eqnarray}
d\tilde{\phi} &=& d\phi + \frac{a}{\Delta}dr \; . 
\end{eqnarray} 
These coordinates are much better suited to numerical evolutions, as
detailed in Ref.~\cite{Krivan1997}. Next, we exploit axisymmetry to
expand $\Psi$ in azimuthal modes:
\begin{eqnarray}
\Psi(t,r,\theta,\tilde{\phi}) & = & \sum_m e^{im\tilde{\phi}}\,r^3\,\phi_m(t,r,\theta)\;.
\end{eqnarray} 
This reduces Eq.\ (\ref{teuk0}) to a set of decoupled
(2+1)-dimensional hyperbolic partial differential equations (PDEs).
We rewrite this system in first-order form by introducing a
momentum-like field,
\begin{equation}
\Pi_m \equiv \partial_t{\phi_m} + \frac {({r}^{2}+{a}^{2})}{\Sigma} \, \partial_{r^*}\phi_m\;,
\end{equation}
where $\Sigma^{2}=(r^2 + a^2)^2 - a^2 \Delta \sin^2 \theta$.  We then
integrate this system using a two-step, 2nd-order Lax-Wendroff
finite-difference method.  Details are presented in
Refs.~\cite{skh07,skh08}. Following Ref.~\cite{Krivan1997}, we set
$\phi_m$ and $\Pi_m$ to zero on the inner and outer radial
boundaries. Symmetries of the spheroidal harmonics are used to
determine the angular boundary conditions: For $m$ even, we have
$\partial_\theta\phi_m =0$ at $\theta = 0,\pi$; for $m$ odd, $\phi_m
=0$ at $\theta = 0,\pi$.

The right-hand side (RHS) of Eq.\ (\ref{teuk0}) is a source term
constructed from the energy-momentum tensor describing a point-like
object moving in the Kerr spacetime.  The expression for $T$ is
lengthy and not particularly illuminating.  For this paper, it is
suffices to point out that $T$ is constructed from Dirac-delta
functions in $r$ and $\theta$, as well as first and second derivatives
of the delta function in these variables.  These terms have
coefficients that are complex functions of the black hole's parameters
and the location of the point-like object. Details and discussion of
how we model the deltas and their derivatives on a numerical grid are
given in Ref.\ {\cite{skh07}}.  The delta functions are sourced at the
location of the point-like object; the source $T$ thus depends on the
trajectory that this body follows in the Kerr spacetime.  In this
analysis, we use a trajectory constructed using the EOB formalism to
specify the small body's location.

One point worth emphasizing is that the source term is scaled by a
factor of $1/\dot t$ [see Eq.\ (2.39) of Ref.~{\cite{skh07}}]; i.e.,
the source is inversely weighted by the rate of change of coordinate
time per unit proper time experienced by the orbiting object.  This
means that the source term ``redshifts away'' as the object approaches
the horizon. As a consequence, when describing a body that falls into
a black hole, the Teukolsky equation (\ref{teuk0}) smoothly
transitions into its homogeneous form, connecting the gravitational
radiation from the last few orbital cycles to the Kerr hole's
quasinormal modes in a very natural way.  The same behavior is seen in
other analyses which model plunging trajectories using black hole
perturbation theory (e.g., Refs.\ {\cite{LoustoPRL:2010,Bernuzzi:2010ty,MinoBrink2008}}).

We implement this numerical scheme with a Fortran code, parallelized
using a standard domain decomposition (on the radial coordinate grid),
and with OpenMPI\footnote{The open source version of MPI, the Message Passing Interface: http://openmpi.org.} enabled message passing.
Good scaling has been observed for several hundred processor cores. In
this analysis, we used 128 processor cores for computing each $m$-mode
for all cases we studied.

\subsection{Waveforms and multipole decomposition}
\label{sec:teukmultipoles}

Far from the black hole, $\psi_4$ is directly related to $h_+$ and
$h_\times$ via
\begin{equation}
\psi_4 = \frac{1}{2}\left(\frac{\partial^2 h_+}{\partial t^2} - i
\frac{\partial^2 h_\times}{\partial t^2}\right)
\equiv \frac{1}{2}\frac{\partial^2 h}{\partial t^2}\,.
\end{equation}
The waveform $h \equiv h_+ - ih_\times$ is then found by integrating
$\psi_4$ twice, choosing constants of integration so that $h \to 0$ at
very late times (long after the system's waves have decayed to zero).

As detailed in Sec.\ {\ref{sec:teukoverview}}, our computation
naturally decomposes the field $\Psi$ (and hence $\psi_4$ and the
waveform $h$) into axial modes with index $m$.  For comparison with
EOB waveforms, it is necessary to further decompose into modes of
spin-weighted spherical harmonics.  Following standard practice, we
define
\begin{eqnarray}
\psi_4 &=& \frac{1}{\cal R}\sum_{\ell,m} C_{\ell m}(t,r)_{-2}Y_{\ell m}(\theta,\phi)\;,
\label{eq:psi4_decomp}\\
h &=& \frac{1}{\cal R}\sum_{\ell,m} h_{\ell m}(t,r)_{-2}Y_{\ell m}(\theta,\phi)\;.
\label{eq:h_decomp}
\end{eqnarray}
In these equations, $_{-2}Y_{\ell m}$ is a spherical harmonic of
spin-weight $-2$.  Defining the inner product
\begin{equation}
\langle Y_{\ell m} | f \rangle = \int d\Omega\, _{-2}Y_{\ell m}^*(\theta,
\phi) f\;,
\end{equation}
(where $^*$ denotes complex conjugation), extracting $C_{\ell m}$ and
$h_{\ell m}$ is simple:
\begin{eqnarray}
C_{\ell m}(t,r) &=& {\cal R}\langle Y_{\ell m} | \psi_4\rangle\;,
\\
h_{\ell m}(t,r) &=& {\cal R}\langle Y_{\ell m} | h\rangle\;.
\end{eqnarray}
The complex wave mode $h_{\ell m}$ can also be obtained from $C_{\ell
  m}$ by integrating twice, again choosing the constants of
integration so that $h_{\ell m} \to 0$ at very late times.

\subsection{Numerical errors}
\label{sec:num-err}

Our numerical solutions are contaminated by two dominant sources of
error: Discretization error due to our finite-difference grid, and
extraction error due to computing $\Psi$ and associated quantities at
finite spatial location rather than at null infinity.

\subsubsection{Discretization error}

As discussed in Ref.\ {\cite{skh08}}, our time-domain Teukolsky solver
is intrinsically second-order accurate.  Since we compute our
solutions on a two-dimensional grid in tortoise radius $r^*$ and angle
$\theta$, we expect our raw numerical output to have errors of order
$(dr^*)^2$, $(d\theta)^2$, and $(dr^* d\theta)$.  We mitigate this
error with a variant of Richardson extrapolation, which we now
describe.

Consider waveforms generated at three resolutions: $h^{(2)}_1$ at
$(dr^{*},d\theta) = (0.064M, 0.2)$; $h^{(2)}_2$ at $(0.032M, 0.1)$;
and $h^{(2)}_3$ at $(0.016M, 0.05)$.  Superscript ``($i$)'' means the
solution is $i$th-order accurate.  We convert from second-order to
third-order accuracy using~\cite{Quarteroni2007}
\begin{eqnarray}
  h^{(3)}_{1.5} &=& h^{(2)}_{1} - \frac{h^{(2)}_{1} - h^{(2)}_{2}}{1-1/n^2}\,,
\nonumber\\
  h^{(3)}_{2.5} &=& h^{(2)}_{2} - \frac{h^{(2)}_{2} - h^{(2)}_{3}}{1-1/n^2}\,.
\label{extrap}
\end{eqnarray}
Here, $n = 2$ is the ratio of grid spacing between the two resolutions.

To estimate the remaining error in this extrapolated solution, we
compare $h^{(3)}_{2.5}$ and $h^{(3)}_{1.5}$.  Let us define
\begin{eqnarray}
\Delta h &=& h^{(3)}_{2.5} - h^{(3)}_{1.5}\;,
\\
h^{(4)} &=& h^{(3)}_{2.5} - \frac{h^{(3)}_{2.5} - h^{(3)}_{2}}{1-1/n^3}\,.
\end{eqnarray}
$h^{(4)}$ is a fourth-order estimate of the Teukolsky solution $h$,
assuming that errors in $h^{(3)}_{2.5,1.5}$ are third order.  Defining
the amplitude $|h|$ and phase $\phi$ as
\begin{equation}
h = |h| e^{i\phi}\;,
\label{eq:ampphasedef}
\end{equation}
the amplitude error $\delta |h|/|h|$ and phase error $\delta\phi$ are
\begin{eqnarray}
\frac{\delta |h|}{|h|} &=& {\rm Re}\left(\frac{\Delta h}{h^{(4)}}\right)\;,
\label{eq:amperr}\\
\delta\phi &=& {\rm Im}\left(\frac{\Delta h}{h^{(4)}}\right)\;.
\label{eq:phaseerr}
\end{eqnarray}

\begin{figure*}
\includegraphics[scale=0.4,bb=150 10 650 510]{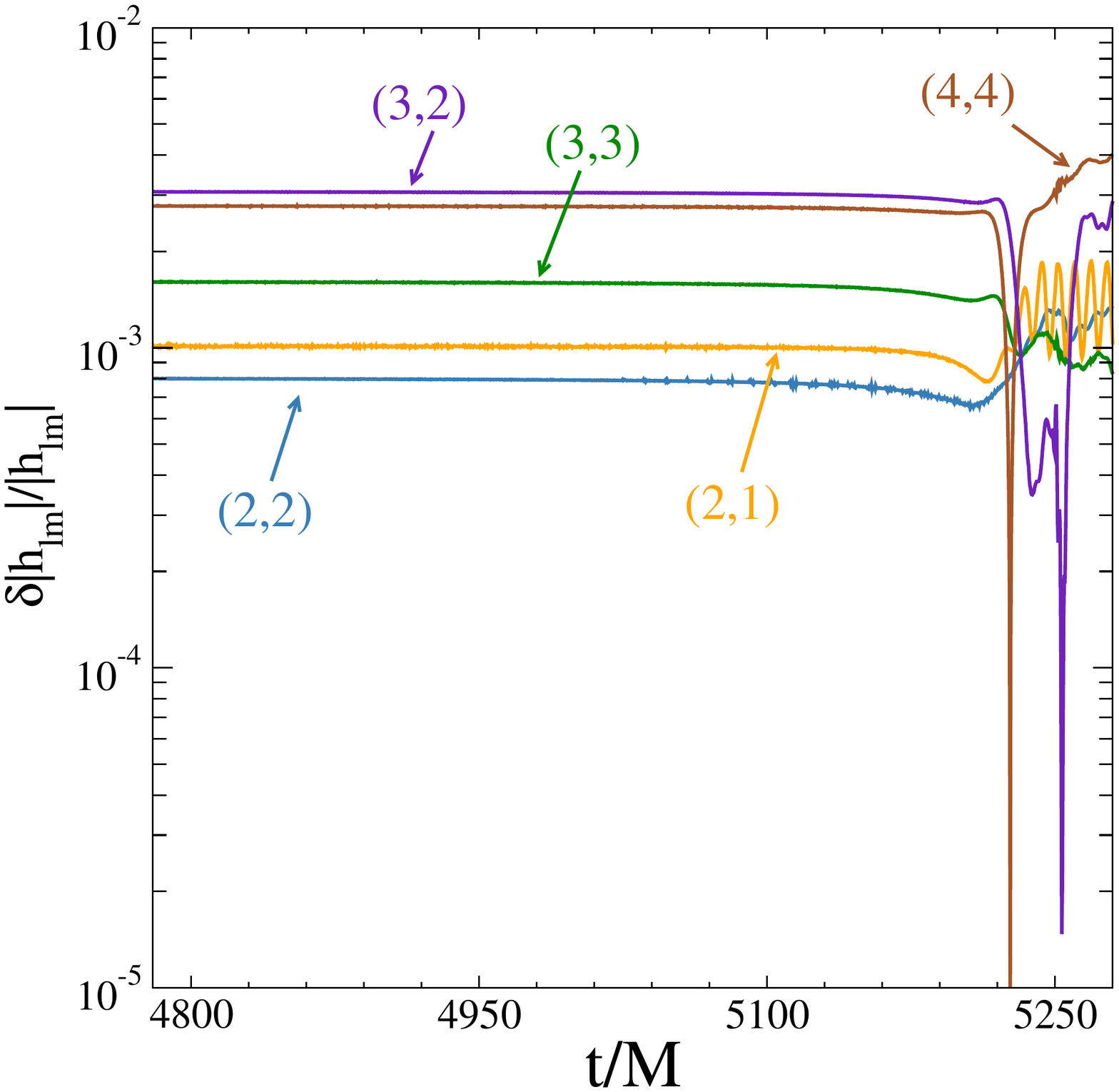}
\includegraphics[scale=0.4,bb=50 10 550 510]{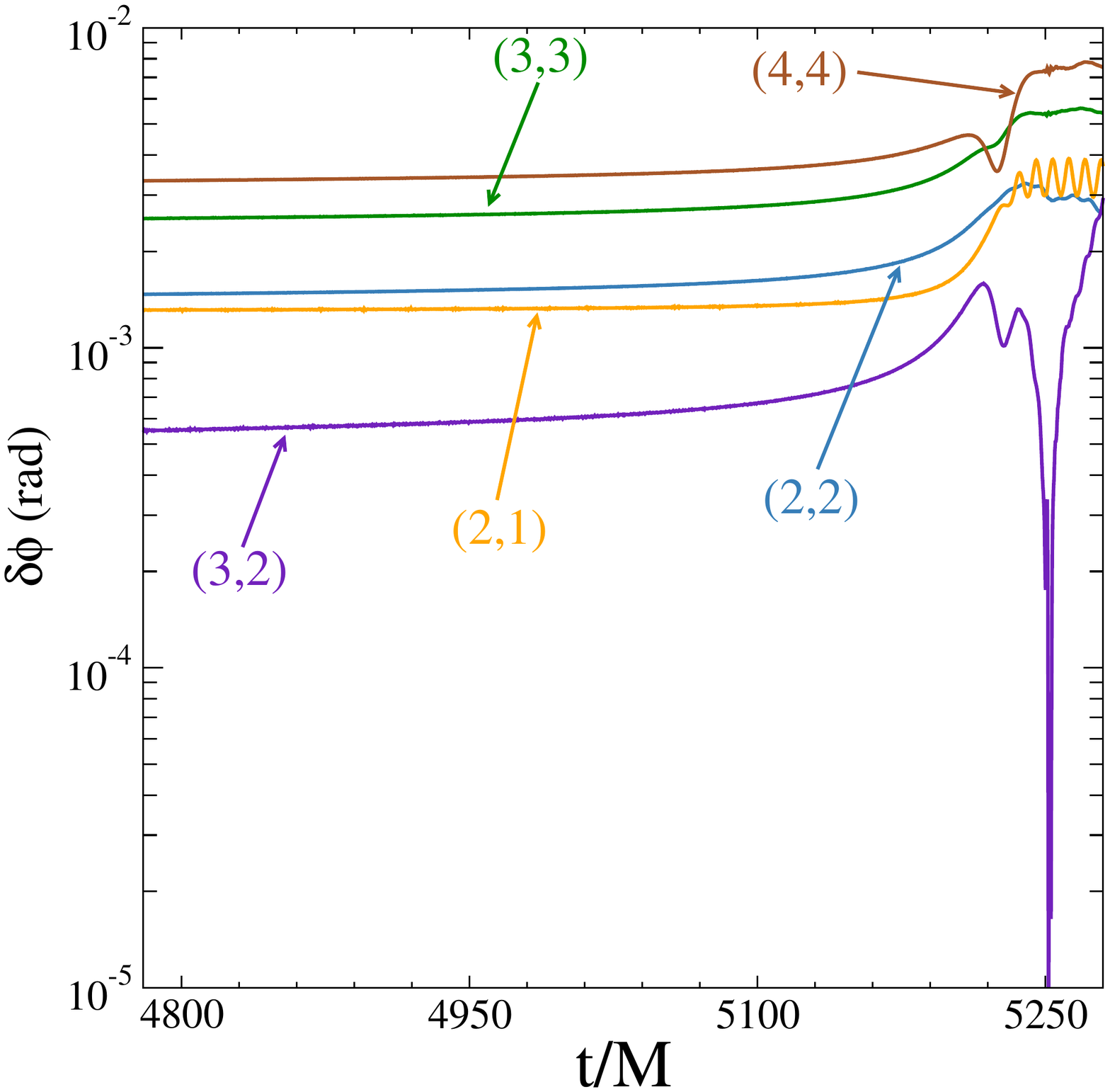} 
\caption{\label{fig:discret_error} Errors in amplitude (left panel)
  and phase (right panel) due to grid discretization for $a = 0$ at
  mass ratio $\mu/M = 10^{-3}$.  These errors are the residual we find
  following the Richardson extrapolation procedure described in the
  text.}
\end{figure*}

Figure \ref{fig:discret_error} shows discretization errors for several
gravitational modes $h_{\ell m}$ extracted at $r^{*} = 950M$.  For
this case, the large black hole is non spinning ($a = 0$).  Amplitude
discretization errors are steady over almost the entire waveform,
until very late times.  In all cases, $\delta |h|/|h| \lesssim \mbox{a
  few}\times10^{-3}$.  Similar behavior is observed for phase errors.
For most modes, $\delta\phi \lesssim \mbox{a few}\times 10^{-3}$
radians over the coalescence.  The highest $(\ell, m)$ modes we
consider approach $10^{-2}$ radian error at the latest times.  Because
higher $(\ell, m)$ modes require higher grid densities to be resolved,
they tend to have larger discretization errors.

\subsubsection{Extraction error}

The code used in the bulk of this analysis extracts $\Psi$ (and
derived quantities such as $\psi_4$ and $h$) at large but finite
radius.  These quantities are more properly extracted at future null
infinity.  Although it has very recently become possible to extract
waveforms at future null infinity (see Refs.\ {\cite{Bernuzzi:2011aj,
    ZenKha2011}}), we did not have this capability when we began this
analysis.  Instead, following Ref.\ {\cite{Boyle2007}}, we extract
waveforms at multiple radii, and fit to a polynomial in $1/r$.
Again defining amplitude $|h|$ and phase $\phi$ using
Eq.\ (\ref{eq:ampphasedef}), we put
\begin{eqnarray}
  |h|(t-r^*,r) &=& |h|_{(0)}(t-r^*) + \sum^{N}_{k=1}\frac{|h|_{(k)}(t-r^*)}{r^k}\;,
\label{amp_poly_fit}\\
  \phi(t-r^*,r) &=& \phi_{(0)}(t-r^*) + \sum^{N}_{k=1}\frac{\phi_{(k)}(t-r^*)}{r^k}\;.
\label{phase_poly_fit}
\end{eqnarray}
The time $t - r^*$ is retarded time, taking into account the finite
speed of propagation to tortoise radius $r^*$; $N$ is the order of the
polynomial fit we choose.  The functions $|h|_{(0)}(t-r^*)$ and
$\phi_{(0)}(t-r^*)$ are the asymptotic amplitudes and phases
describing the waves at future null infinity.

We extract waveforms at radii $r = 150M$, $350M$, $550M$, $750M$ and
$950M$.  We then perform non-linear, least-squares fits for $|h|_{(k)}$
and $\phi_{(k)}$ using the Levenberg-Marquardt method \cite{numrec_c}
to find the asymptotic waveform amplitudes and phases.  Following
Ref.\ {\cite{Boyle2007}}, we use $N = 3$ for the order of our fit, and
estimate errors by comparing the fits for $N = 3$ and $N = 2$.

\begin{figure*}
\includegraphics[width=8.5cm,clip=true]{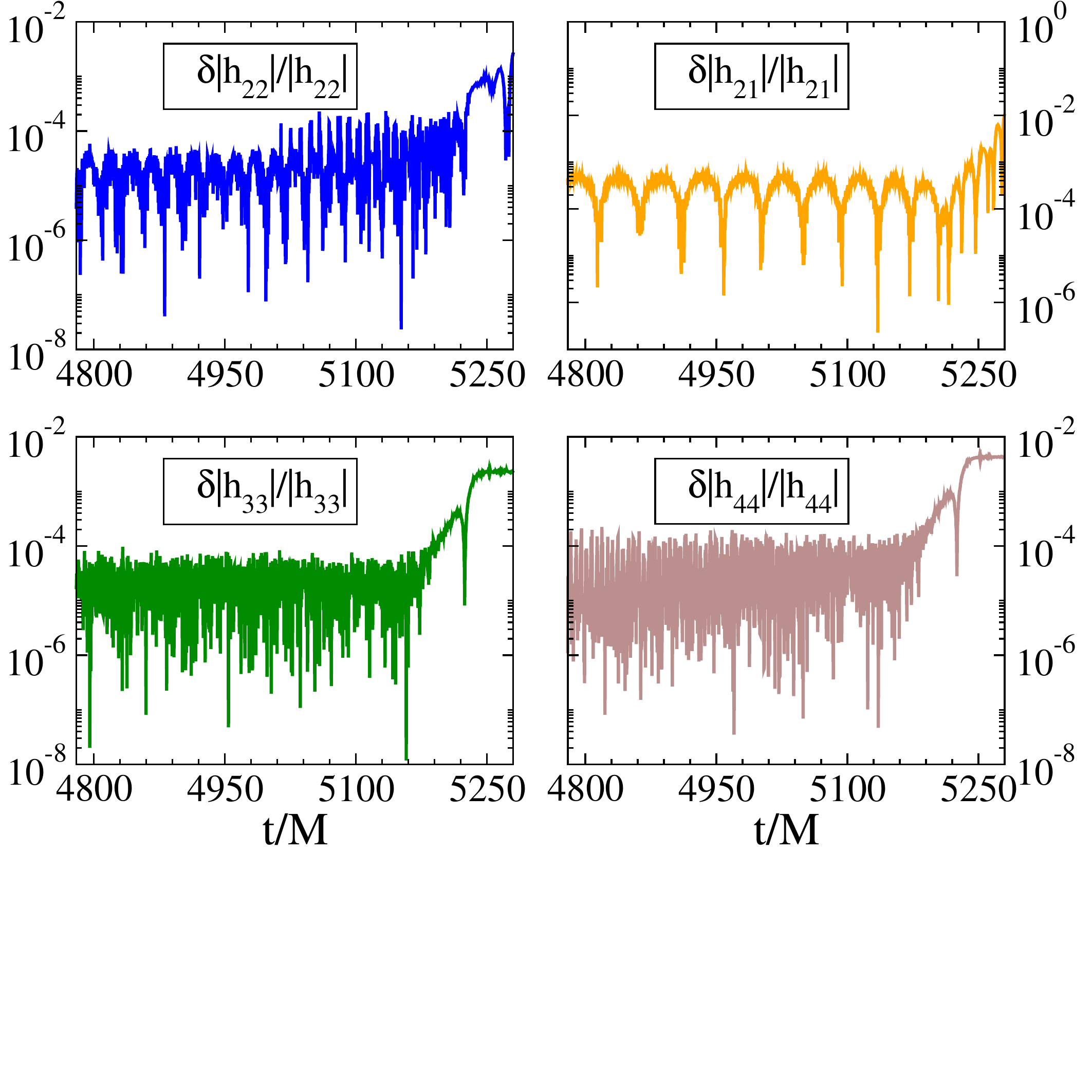} 
\includegraphics[width=8.6cm,clip=true]{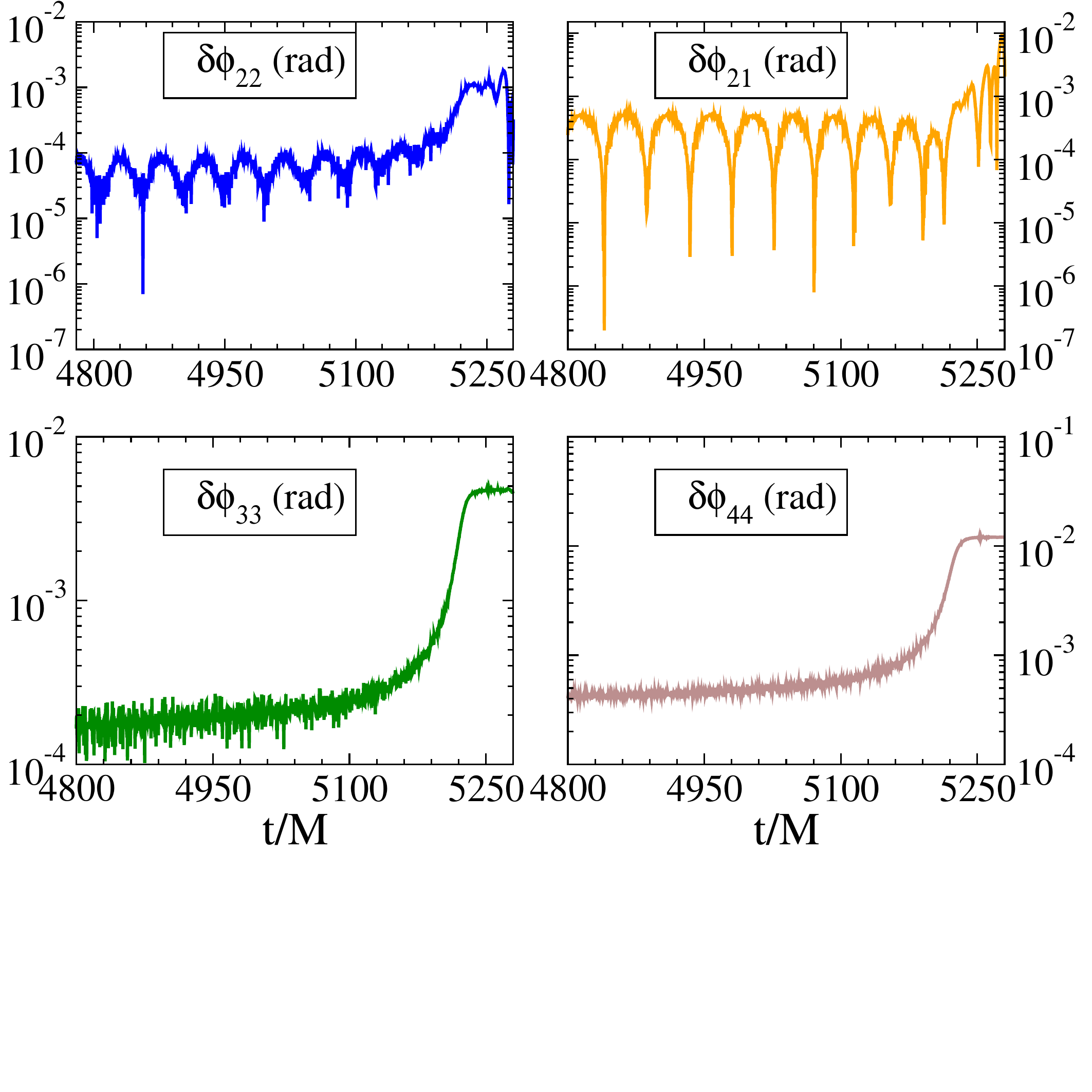} 
\vspace{-2cm}
\caption{\label{fig:extrap_error} Errors in amplitude (left panel) and
  phase (right panel) following extrapolation to infinity for the
  non-spinning case at mass ratio $\mu/M = 10^{-3}$.}
\end{figure*}

Figure \ref{fig:extrap_error} shows the extrapolation errors we find
for the same case shown in Fig.\ \ref{fig:discret_error}.  For most of
the evolution, extrapolation errors are smaller than discretization
errors.  In particular, the amplitude errors are at or below $10^{-3}$
for most of the coalescence; phase errors are at or below $10^{-3}$
radians.  Both phase and amplitude errors grow to roughly $10^{-2}$
very late in the evolution.   
Note that the largest errors in $\psi_4$ come at the latest times, when the waves have largely decayed away. In other
words, the largest errors occur when the waves are weakest. Because we compute $\psi_4$ and then infer amplitude and phase, both amplitude and phase are affected in roughly equal measure by these late time errors [see Eqs. (\ref{eq:amperr}) and (\ref{eq:phaseerr})].
Our
numerical errors appear to be of similar size to error estimates seen
in related analyses (e.g., Ref.\ \cite{Bernuzzi:2011aj}).

Finally, it is worth noting that, thanks to the hyperboloidal layer
method introduced to time-domain black hole perturbation theory in
Refs.\ \cite{Bernuzzi:2011aj, ZenKha2011}, it will not
be necessary to perform this extrapolation in future work.  The codes
will, to very good accuracy, compute the waveform directly at future
null infinity.  Although this advance did not come in time for the
bulk of our present analysis, we have used it to check our error
estimates in several cases.  We find that our total numerical error
estimates (discretization plus extrapolation error, combined in
quadrature) is similar to the errors we compute using the
hyperboloidal layer method\footnote{It is worth emphasizing that codes
  which use the hyperboloidal layer method are much faster than those
  which use the extrapolation described here; we find a speedup of
  roughly ten (for the scale of the evolutions performed 
  in the context of this work).  Although it is gratifying that these
  extrapolations reliably improve our numerical accuracy, the
  substantial speed-up means that upgrading our method is worthwhile
  for future work.}.  This gives us confidence that our error
estimates are reliable.

\subsection{Comparing time-domain and frequency-domain Teukolsky codes}
\label{sec:FD-TD}

As a further check on the accuracy of our numerical Teukolsky-based
waveforms, we compare time-domain (TD) waveforms computed using the
techniques described here with frequency-domain (FD) waveforms
\cite{h00,dh06}.  Since we only calibrate the higher-order modes in
the EOB model for $a = 0$, we focus on that case here.  We expect our
conclusions to be similar for spinning cases since we use the same
procedure to estimate errors in that case.

As described in Secs.\ {\ref{sec:intro}} and {\ref{sec:teukoverview}},
in our analysis the source term for the TD waveforms
[cf.\ Eq.\ (\ref{teuk0})] depends on the EOB inspiral and plunge
trajectory.  For FD waveforms, by contrast, the source is built from a
purely geodesic trajectory.  This is because the FD code uses the
existence of discrete orbital frequencies.  For this analysis, we
specialize further to circular-orbit equatorial geodesics, but allow
these geodesics to evolve adiabatically using FD Teukolsky fluxes, as
described in Ref.\ {\cite{h01}}.  Previous work has shown that a
self-consistent adiabatic evolution implemented with our FD code is in
excellent agreement with the EOB model during the inspiral
{\cite{Yunes:2009ef}}, and so it makes sense to compare TD and FD
waveforms during this phase of the coalescence.  It is also worth
noting that FD waveforms can generally be computed to near machine
accuracy using spectral techniques~\cite{ft05}.  The only limitation
on their accuracy is truncation of the (formally infinite) sums over
multipoles and frequency harmonics.  We can thus safely assume that
the difference between TD and FD waveforms is only due to errors in
the TD waves.

To perform this comparison, we align the $\ell=m=2$ TD and FD
waveforms by introducing time and phase shifts $\Delta t$ and $\Delta
\phi$ that minimize the gravitational phase difference at low
frequencies. More specifically, we choose $\Delta t$ and $\Delta \phi$
in order to minimize
\begin{equation}\label{deltaphiSq}
\int_{t_1}^{t_2} \left[\phi^{\rm FD}_{\rm 22}(t)-\phi^{\rm TD}_{\rm
    22}(t+\Delta t)+\Delta \phi\right]^2 {\rm d}t\,,
\end{equation}
where $t_1$ and $t_2$ are separated by $1000M$ and correspond to $M
\omega_{\rm 22}\approx 0.108$ and a $M \omega_{\rm 22}\approx 0.111$,
respectively.  This low-frequency alignment is necessary for three
reasons.  First, the time coordinate of the TD waveform includes the
effect of the extraction radii of the data used for the extrapolation;
the FD waveforms are truly extracted at future null infinity.  Second,
the initial phases of the TD and FD trajectories are not necessarily
the same, which introduces a phase offset between the two models.
Third and last, as discussed in detail in Ref.\ \cite{skh07}, TD
waveforms include an initial burst of ``junk'' radiation, which must
be discarded.  During that burst, the TD and FD trajectories may
accumulate a small phase difference.  We have found that small changes
to $t_1$ and $t_2$ do not significantly affect the alignment.

Once $\Delta t$ and $\Delta \phi$ are fixed, we have no freedom to
introduce further time or phase shifts for the other modes. For
instance, the difference between the FD and TD phases for the mode
$(\ell,m)$ is
\begin{equation}\label{phasediff}
\delta\phi^{\rm FD-TD}_{\ell m}\equiv\left\vert\phi^{\rm FD}_{\ell m}(t)-\phi^{\rm
  TD}_{\ell m}(t+\Delta t)+m\frac{\Delta \phi}{2}\right\vert\,.
\end{equation}
The fractional amplitude difference is
\begin{equation}\label{ampdiff}
\frac{\delta |h|^{\rm FD-TD}_{\ell m}}{|h|_{\ell m}}\equiv\left\vert
\frac{|h|^{\rm FD}_{\ell m}(t)}{|h|^{\rm TD}_{\ell m}(t+\Delta
  t)}-1\right\vert\,.
\end{equation}
The $\Delta t$ and $\Delta \phi$ used here are the ones that minimize
\eqref{deltaphiSq}.

\begin{table}
\centering 
\begin{tabular}{c| c| c| c| c} 
\hline\hline 
$(\ell, m)$ & $\delta\phi^{\rm FD-TD}_{\ell m}$ & $\delta\phi^{\rm TD}_{\ell m}$
& $\displaystyle\frac{\delta |h|^{\rm FD-TD}_{\ell m}}{|h|_{\ell m}}$
& $\displaystyle\frac{\delta |h|^{\rm TD}_{\ell m}}{|h|_{\ell m}}$  \T \B\\  
\hline 
 (2,2) & 7.71$\times 10^{-4}$ & 1.27$\times 10^{-3}$ & 1.22$\times 10^{-4}$ & 8.20$\times 10^{-4}$ \\
 (3,3) & 1.18$\times 10^{-3}$ & 2.22$\times 10^{-3}$ & 1.95$\times 10^{-4}$ & 1.64$\times 10^{-3}$\\
 (2,1) & 4.05$\times 10^{-4}$ & 1.28$\times 10^{-3}$ & 2.55$\times 10^{-4}$ & 1.03$\times 10^{-3}$\\
 (4,4) & 1.58$\times 10^{-3}$ & 2.94$\times 10^{-3}$ & 2.80$\times 10^{-4}$ & 2.80$\times 10^{-3}$ \\
 (3,2) & 7.71$\times 10^{-4}$ & 3.92$\times 10^{-4}$ & 4.09$\times 10^{-4}$ & 3.13$\times 10^{-3}$\\
\hline\hline
\end{tabular}
\caption{The phase difference and fractional amplitude difference for
  various modes, averaged over the time interval $t_2 - t_1$ [see
    Eq.\ (\ref{deltaphiSq})].  We compare the amplitude and phase
  error found by comparing TD and FD waveforms (columns 2 and 4) with
  the TD errors we estimate using the techniques discussed in
  Sec.\ {\ref{sec:num-err}}.  In all cases but one [phase error for
    the $(3,2)$ mode], our numerical error estimates are larger than
  those we find comparing the two calculations; in that single
  discrepant case, the errors themselves are particularly small.  This is further
  evidence that our numerical error estimates are reliable.}
\label{FDvsTD}
\end{table}

Table~\ref{FDvsTD} compares $\delta\phi^{\rm FD-TD}_{\ell m}$ and
$\delta |h|^{\rm FD-TD}_{\ell m}/|h|_{\ell m}$ with the errors computed using the
techniques described in Sec.\ {\ref{sec:num-err}}.  In particular, we
examine the averages of $\delta\phi^{\rm FD-TD}_{\ell m}$ and $\delta
|h|^{\rm FD-TD}_{\ell m}/|h|_{\ell m}$ over the alignment interval
$(t_1,t_2)$, and compare them to the averages over the same interval
of the TD numerical errors discussed in the previous section.  For
this comparison, we average the sum (in quadrature)  of discretization
and extrapolation errors.   We see that the difference between TD and
FD is always within the TD numerical errors, except for the $\ell=3$,
$m=2$ mode.  This mode is among the weakest of those that we show in
Table~\ref{FDvsTD}, which makes the extraction procedure described in
the previous section considerably more difficult.

\subsection{Characteristics of time-domain Teukolsky merger waveforms}
\label{sec:characteristics}

\begin{figure*}
\includegraphics[scale=0.4,bb=150 10 650 510]{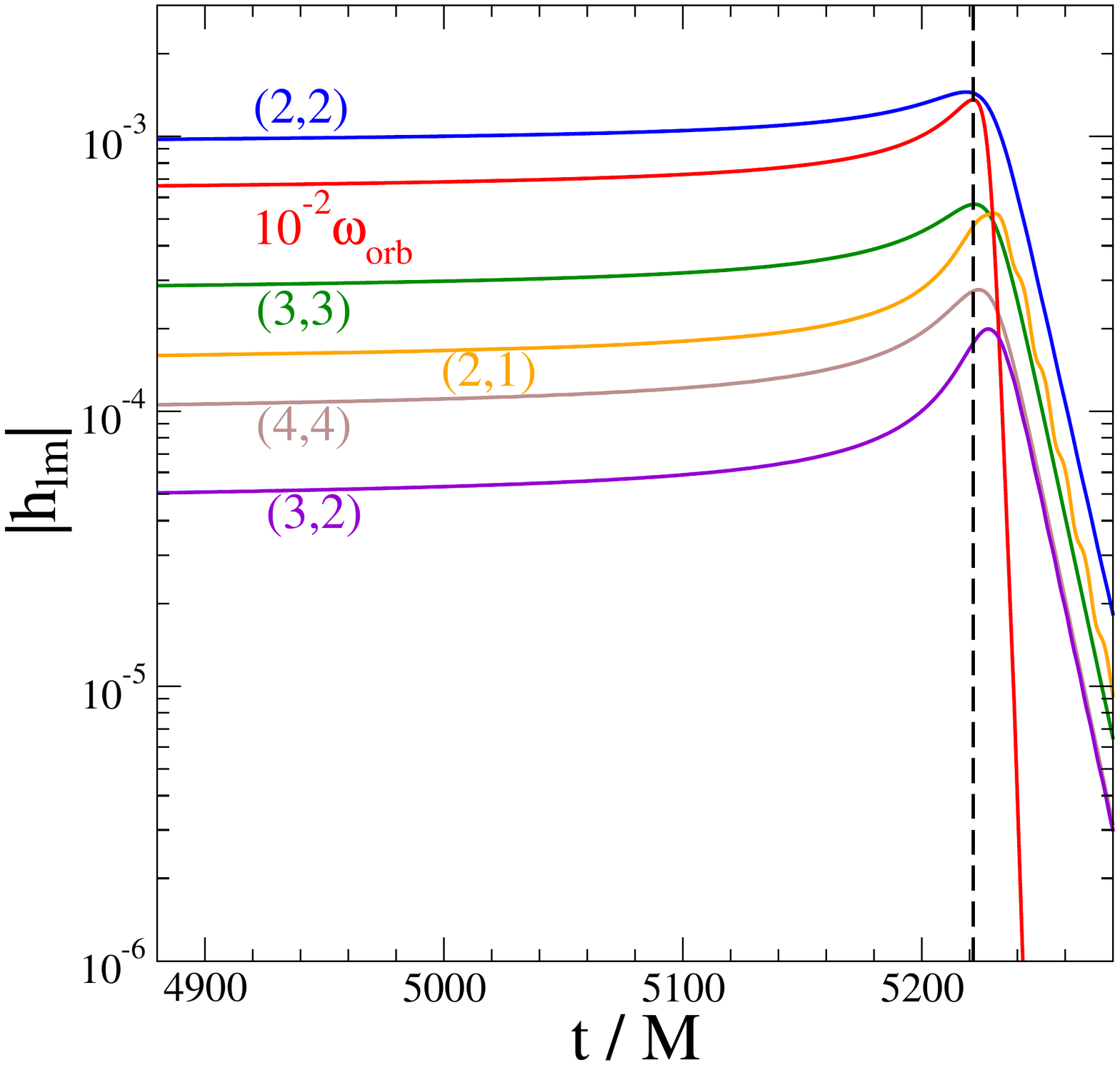}
\includegraphics[scale=0.4,bb=50 10 550 510]{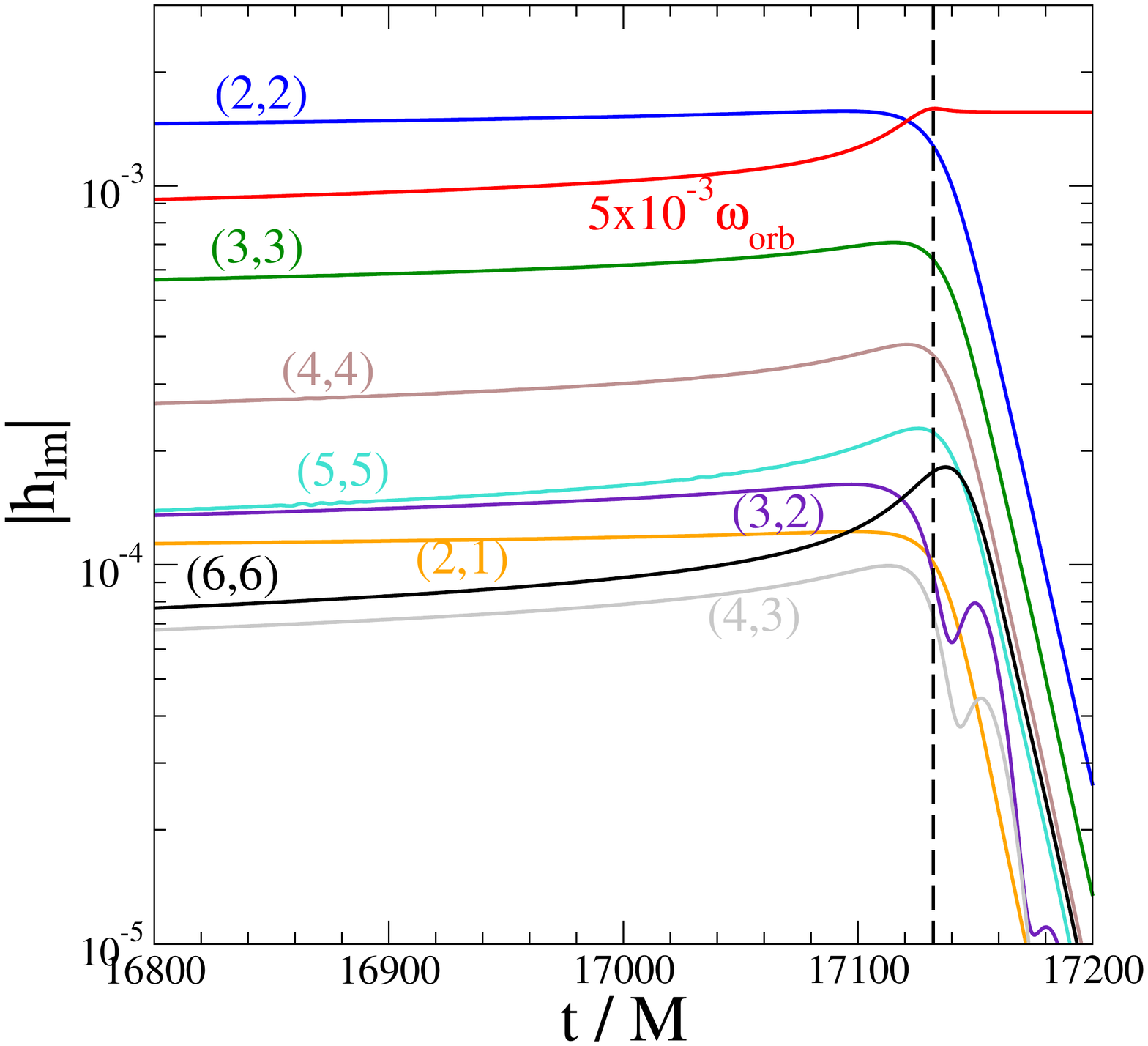}
\vspace{-0.3cm}
\caption{\label{fig:modesampl} Amplitude of the dominant modes during
  plunge, merger and ringdown for $a = 0$ (left panel) and $a/M = 0.9$
  (right panel).  We also show orbital frequency, scaled to fit on the
  plot. As expected, the orbital frequency asymptotes to the horizon's angular velocity at late times, because the frame dragging
locks the particle's motion to that of the horizon. The vertical dashed line in the two panels marks the position
  of the peak of the orbital frequency.}
\end{figure*}

We now turn to the waveforms produced by the TD Teukolsky analysis and
their general characteristics.  Figure \ref{fig:modesampl} examines
the behavior of the dominant TD modes [$(\ell,m)=(2,2), (3,3), (4,4),
  (2,1), (3,2)$] during plunge, merger and ringdown.  We also show the
orbital frequency of the EOB trajectory used to produce the TD data.

For the non-spinning case (left panel), the peak of the $(2,2)$
amplitude comes slightly earlier than the peak in orbital frequency,
while higher-order modes peak later.   A summary of the time difference
$t_{\rm peak}^{\rm \ell m}-t_{\rm peak}^{\Omega}$ between the peak of
the Teukolsky mode amplitude and that of the EOB orbital frequency is
shown in Table~\ref{tab:delay_nonspinning}.  This difference can be as
large as $6.25 M$ for the $(3,2)$ mode, and even $8.82 M$ for the
$(2,1)$ mode.

\begin{table}
\centering 
\begin{tabular}{c| c} 
\hline\hline 
$(\ell, m)$ & $(t_{\rm peak}^{\rm \ell m}-t_{\rm peak}^{\rm  \Omega})/M$ \T \B\\
\hline 
 (2,2) & -2.99\\
 (3,3) & 0.52\\
 (4,4) & 2.26\\
 (2,1) & 8.82\\
 (3,2) & 6.25\\
\hline\hline
\end{tabular}
\caption{The time difference $(t_{\rm peak}^{\ell m}-t_{\rm
    peak}^{\Omega})/M$ between the peak of the Teukolsky modes'
  amplitude and of the orbital frequency, for $a/M=0$.}
\label{tab:delay_nonspinning}
\end{table}

The situation is different in the spinning case.  A trend we see is
that as the spin $a$ grows positive, higher-order modes become
progressively more important.  In the right-hand panel of
Fig.\ {\ref{fig:modesampl}}, we show the amplitude of the eight
strongest modes for $a/M = 0.9$.  As noticed in Ref.~\cite{Pan2010hz},
modes with $\ell = m$ tend to increase more during the plunge.  For
example, the $(5,5)$ mode is smaller than the $(3,2)$ mode during
inspiral, but becomes larger during the plunge. The $(6,6)$ mode also
grows quickly during the plunge (which roughly ends at the time of
the peak of the orbital frequency).  This behavior is not surprising,
in fact should become more and more pronounced as the spin $a/M \to 1$;
modes with large multipole moments become as important as
low-$\ell$ modes in that limit, at least for quasi-circular
orbits~\cite{finn_thorne,naked_sings}.

\begin{table}
\centering 
\begin{tabular}{c|c|c} 
\hline\hline
$a/M$ & $(t_{\rm peak}^{\rm 22}-t_{\rm peak}^{\Omega})/M$ & $M {\Omega}_{\rm max}$\T \B\\
\hline 
 -0.9  & 1.60 &  0.090\\
 -0.5 & -0.08 &   0.106 \\
  0 & -2.99 & 0.136 \\
 0.5  & -7.22 &  0.193 \\
 0.7  & -12.77 &  0.236 \\
0.9  & -39.09 &  0.320 \\
\hline\hline
\end{tabular}
\caption{The time difference $(t_{\rm peak}^{22}-t_{\rm
    peak}^{\Omega})/M$ between the peak of the Teukolsky $(2,2)$
  amplitude of the orbital frequency, for various values of the spin
  $a/M$. Also shown is the value of the orbital frequency at that
  peak, $M {\Omega}_{\rm max}$.}
\label{tab:delay_spinning}
\end{table}

Table \ref{tab:delay_spinning} shows the time difference $(t_{\rm
  peak}^{\rm 22}-t_{\rm peak}^{\Omega})/M$ between the peak of the
$(2,2)$ amplitude and of the orbital frequency for the values of spin
that we consider in this paper.  As the spin grows, the orbital
frequency peaks later and later relative to the peak of the $(2,2)$
amplitude.  This has important implications for modeling the EOB
merger-ringdown waveform, as we discuss in detail later in the paper.

\begin{figure}
\includegraphics[width=8.5cm,clip=true]{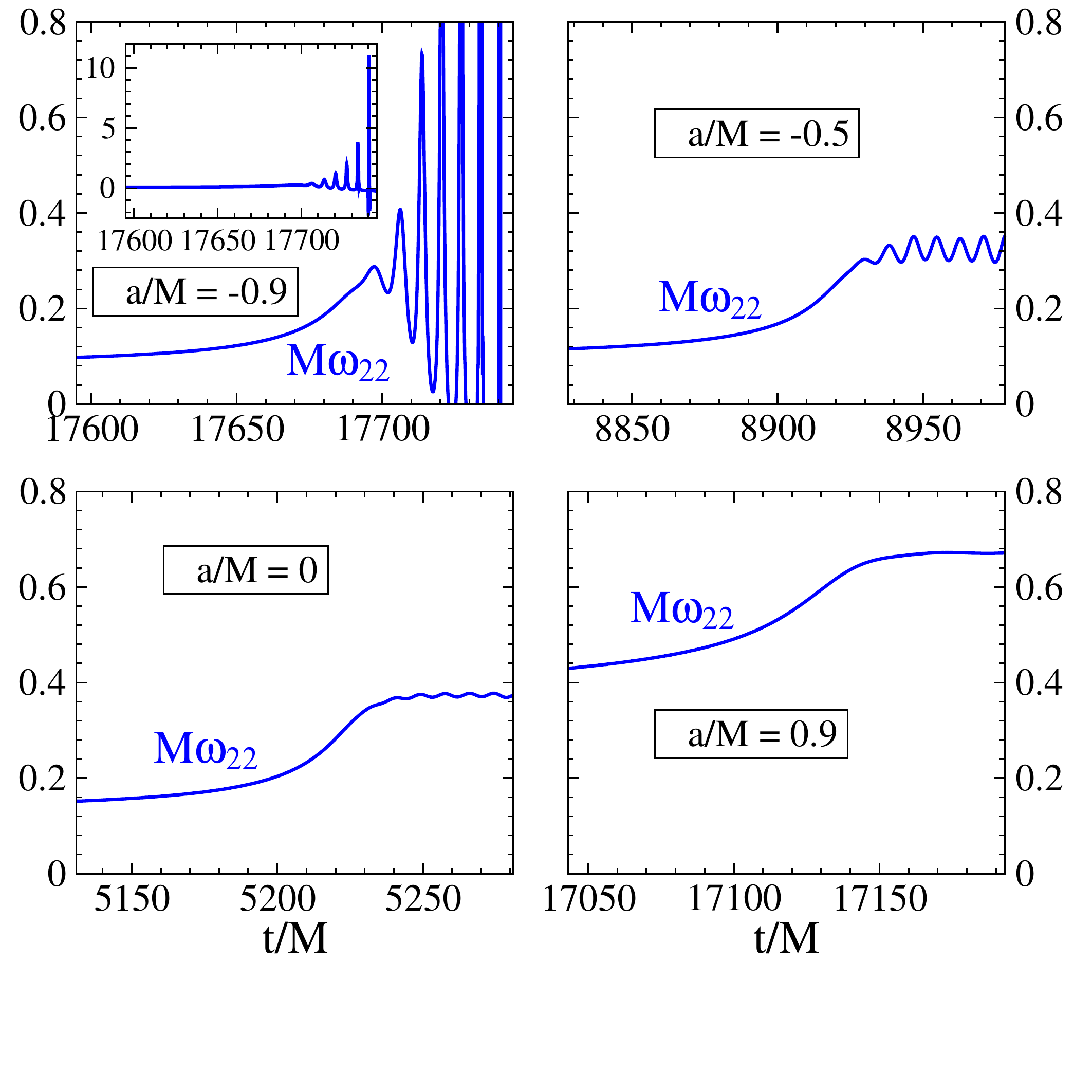} 
\vspace{-1cm}
\caption{\label{fig:freq22} The evolution of wave frequency for the
  $(2,2)$ mode for several spins.  Notice that the strong late
  oscillations (during ringdown), which are particularly prominent as
  the spin decreases towards negative values; they are especially
  large for $a/M = -0.9$.  These oscillations are physical, and arise
  from superposition of the $(2,\pm2)$ QNMs.}
\end{figure}

Another interesting feature of the Teukolsky waveforms that we find is
shown in Fig.~\ref{fig:freq22}.  This figure shows the
gravitational-wave frequency (defined as the time derivative of the
phase) for the $(2,2)$ mode during plunge, merger, and ringdown, for
several spin values.  Notice the strong oscillations seen at late
times (during the final ringdown) for spins $a\lesssim0$.  These
oscillations grow as the spin decreases, and become very large for
$a/M=-0.9$. We have verified that these oscillations (even in the
$a/M=-0.9$ case) are \textit{not} numerical artifacts.  We have found
that they are insensitive to numerical resolution and floating-point
precision, that they also appear in the context of other plunging
retrograde trajectories.  We find that they are due to a superposition
of the dominant $(2,2)$ QNM with the $(2,-2)$ QNM, which is also
excited during the plunge {\cite{Krivan1997,Bernuzzi:2010ty}}.

In order to extract the relative amplitudes of these modes, we fit the
gravitational-wave frequency in a region where the higher overtones of
the $(2,\pm2)$ modes have decayed away and the waveform can be
described by
\begin{equation}
\label{hosc}
h(t)= h_0 (e^{-i \sigma_{220}\,t} + \bar{\alpha}\,e^{i \sigma^*_{2-20}\,t+\bar{\beta}})\,,
\end{equation}
where $\sigma_{2\pm20} = \omega_{2\pm20}-i/\tau_{2\pm20}$ are the
complex QNM frequencies [with overtone $n=0$, as introduced in
  Eq.~(\ref{RD})].  The complex parameter $h_0$ and real parameters
$\bar{\alpha}$, $\bar{\beta}$ are left unspecified. From
Eq.~(\ref{hosc}) one can then calculate the frequency as $\Re[ -i
  \dot{h}(t)/h(t)]$.  Because $h_0$ cancels in this expression, we are
left with the parameters $\bar{\alpha}$ and $\bar{\beta}$, which can
be determined by numerical fitting {\cite{Bernuzzi:2010ty}}.  We find
that the relative excitation $\bar{\alpha}$ of the $(2,-2)$ modes goes
from $\bar{\alpha}\approx0.005$ for $a/M=0$, to
$\bar{\alpha}\approx0.03$ for $a/M=-0.5$ and $\bar{\alpha}\approx0.46$
for $a/M=-0.9$.  This is nicely in accord with the growing strength of
the oscillations as $a/M \to -1$ that is seen in
Fig.\ {\ref{fig:freq22}}.

A possible reason why the $(2,-2)$ QNM is strongly excited for large
negative spins can be understood by examining the particle's
trajectory.  When $a<0$, the spin angular momentum is oppositely
directed from the orbital angular momentum.  During the inspiral, when
the orbit is very wide, the orbit's angular velocity is opposite to
the sense in which the horizon rotates.  At late times (during the
final plunge), the particle's motion becomes locked to the horizon by
frame dragging.  The particle's angular velocity thus flips sign at
some point during the plunge when $a < 0$.  This change in angular
velocity is most pronounced for large negative spins, since the
difference between the frequency at the innermost stable circular
orbit (ISCO) and at the event horizon is largest for large negative
$a$.

\begin{figure}
\includegraphics[width=8cm,clip=true]{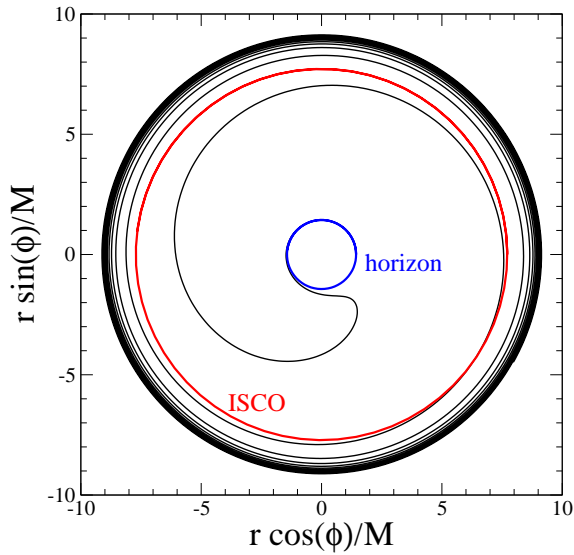} 
\caption{\label{fig:orbit} The EOB equatorial orbit used to produce
  the Teukolsky waveforms for $a/M = -0.9$, focusing on the transition
  between the inspiral and the plunge.  As shown here, the Kerr black
  hole rotates clockwise. The red external circle is the ISCO; the
  blue internal one is the event horizon.  The particle initially
  moves anti-clockwise on quasi-circular orbits, but flips to
  clockwise motion during the plunge as its angular motion becomes
  locked to the horizon's motion.}
\end{figure}

Figure \ref{fig:orbit} shows, as an example, the EOB trajectory we
used to produce the $a/M=-0.9$ Teukolsky waveforms.  As viewed here,
the horizon rotates in the clockwise sense.  After the anti-clockwise
inspiral, the particle plunges and its angular velocity flips sign
before the particles settles on a quasi-circular orbit with $r\to r_+$
and $\Omega\to \Omega_+ = a/2Mr_+$ as $t\to+\infty$ (where $r_+ = M +
\sqrt{M^2 - a^2}$ is the coordinate radius of the event horizon).
This behavior leads us to conjecture that the $(2,-2)$ QNM is excited
by the last part of the plunge, when the particle is corotating with
the black hole.  The $(2,2)$ QNM is excited by the final inspiral and
initial plunge, when the particle is counter-rotating relative to the
black hole.  When $a>0$ the particle's motion is always co-rotating
with the black hole, both during inspiral and through the plunge.
This conjecture thus explains why oscillations in the ringdown
frequency are much less significant for $a\gtrsim0$ and seem to
disappear when $a/M\approx1$ (see Fig.~\ref{fig:freq22}).

\section{Comparison of the EOB model with the Teukolsky time-domain waveforms}
\label{sec:results}

In this Section, we present the main results of this paper, comparing
the EOB waveforms for binary coalescence with waveforms calculated
using the time-domain Teukolsky equation tools described in the
previous section.  We begin by comparing Teukolsky waveforms (for $a =
0$) with an EOB model that has been calibrated for the comparable-mass
case (Sec.\ {\ref{sec:results_original}}).  The agreement is good for
some modes [$(2,2)$ and $(3,3)$], but is much less good for others
[$(2,1)$ and $(4,4)$].  We nail down the reason for this disagreement,
recalibrate the EOB model, and show much better agreement in the
comparison in Sec.\ {\ref{sec:results_improved}}.  We then consider $a
\ne 0$.  Focusing on the $(2,2)$ mode, we compare Teukolsky and EOB
waveforms for a range of spins in Sec.\ \ref{sec:results_kerr_eob}.

We stress that all the comparisons that we present in this paper have
been performed between EOB and Teukolsky waveforms produced with the \textit{same} 
EOB trajectory. For example, in order to recalibrate the EOB model,
we start with a reasonable EOB trajectory, we feed that to the Teukolsky code, and compare the resulting
Teukolsky waveforms to the EOB waveforms. If the waveforms do not agree, we modify the EOB trajectory so that the EOB waveforms agree with the Teukolsky waveforms produced with the old EOB trajectory, and then feed the new EOB trajectory to the Teukolsky code.
We then iterate until this procedure has converged.

\subsection{Comparison of comparable-mass EOB waveforms and Teukolsky waveforms for $a=0$}
\label{sec:results_original}

Reference \cite{Pan:2011gk} presents an EOB model calibrated to
numerical-relativity simulations of non-spinning black-hole binaries
with mass ratios $m_2/m_1=1, 1/2, 1/3, 1/4$ and $1/6$.  This model
achieves very good agreement between the phase and amplitude of the
EOB and numerical-relativity waveforms; see Secs.\ II and III of
Ref.~\cite{Pan:2011gk} for details.  As background for the comparison
we will make to the Teukolsky waveform, we briefly discuss how the EOB
inspiral-plunge waveform was built, and how the merger-ringdown
waveform was attached to build the full waveform

\begin{figure}
\includegraphics[width=8cm,clip=true]{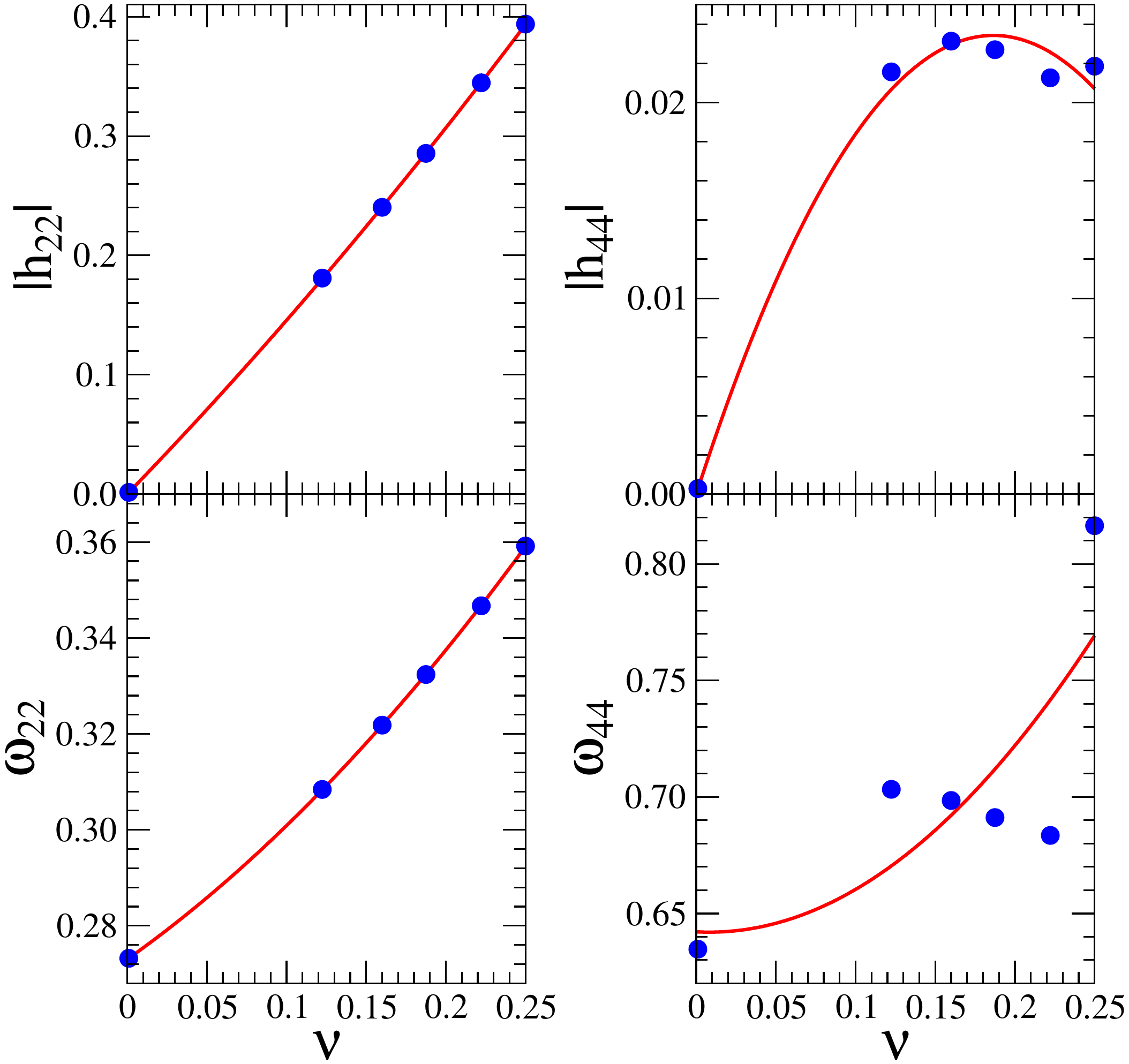} 
\caption{\label{fig:peakvaluefit} The amplitude $h$ (top panels) and
  gravitational-wave frequency $\omega$ (bottom panels) when the
  $(2,2)$ (left) and $(4,4)$ (right) modes reach their peak.  Circles
  at $\nu \ge 0.12$ denote data points extracted from the
  numerical-relativity simulations; the left-most points at $\nu =
  10^{-3}$ are data extracted with the Teukolsky code.  The solid
  lines are quadratic fits to the data points.}
\end{figure}

For each mode, Ref.~\cite{Pan:2011gk} set $a_4^{h_{\ell
    m}}=a_5^{h_{\ell m}}=b_3^{h_{\ell m}}=0$ in Eq.~(\ref{Nlm}), and
fixed the remaining coefficients $a_i^{h_{\ell m}}$ [$i \in (1,2,3)$]
and $b_i^{h_{\ell m}}$ [$i \in (1,2)$] by imposing the following five
conditions:

\begin{figure*}
\includegraphics[width=8cm,clip=true]{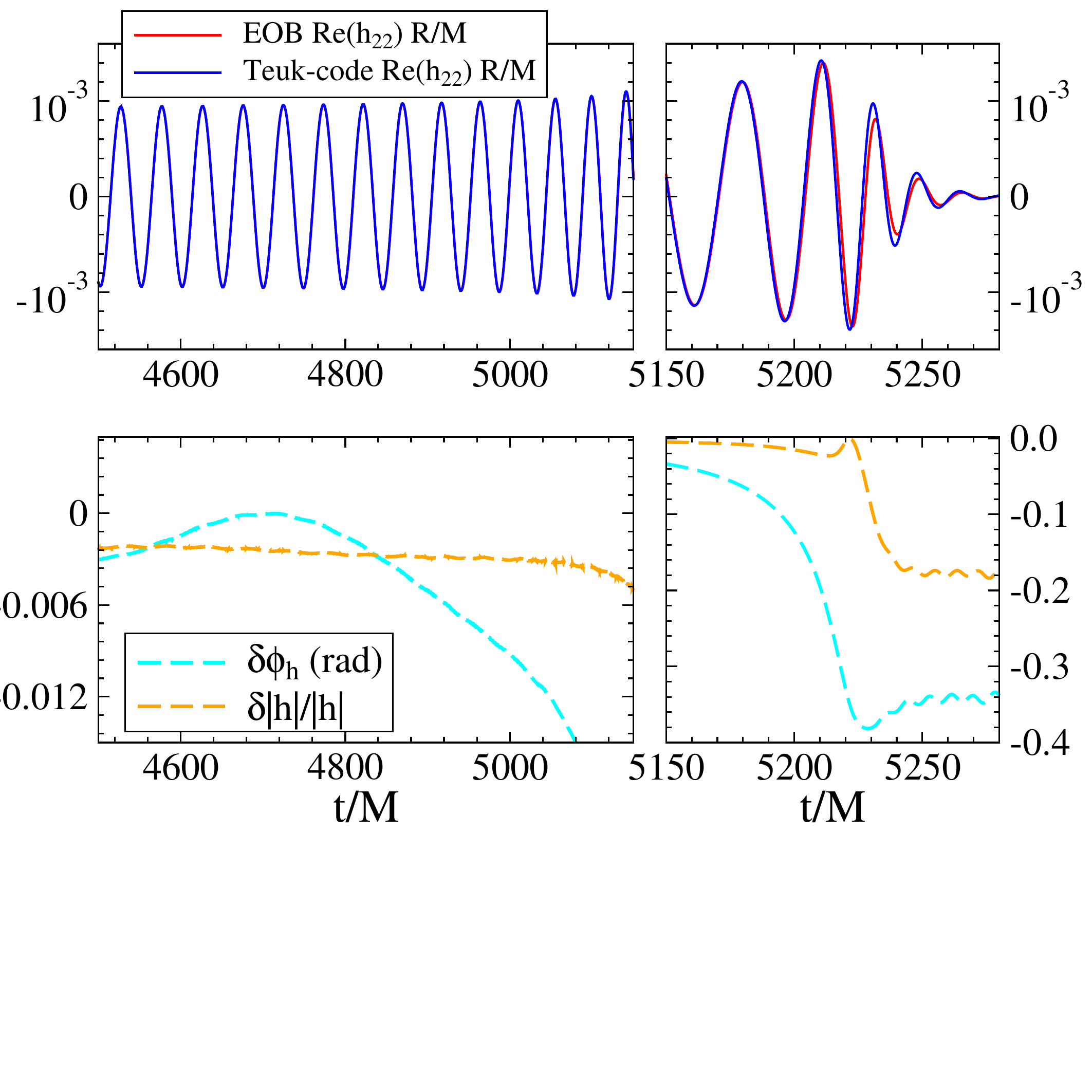} \qquad \qquad
\includegraphics[width=8cm,clip=true]{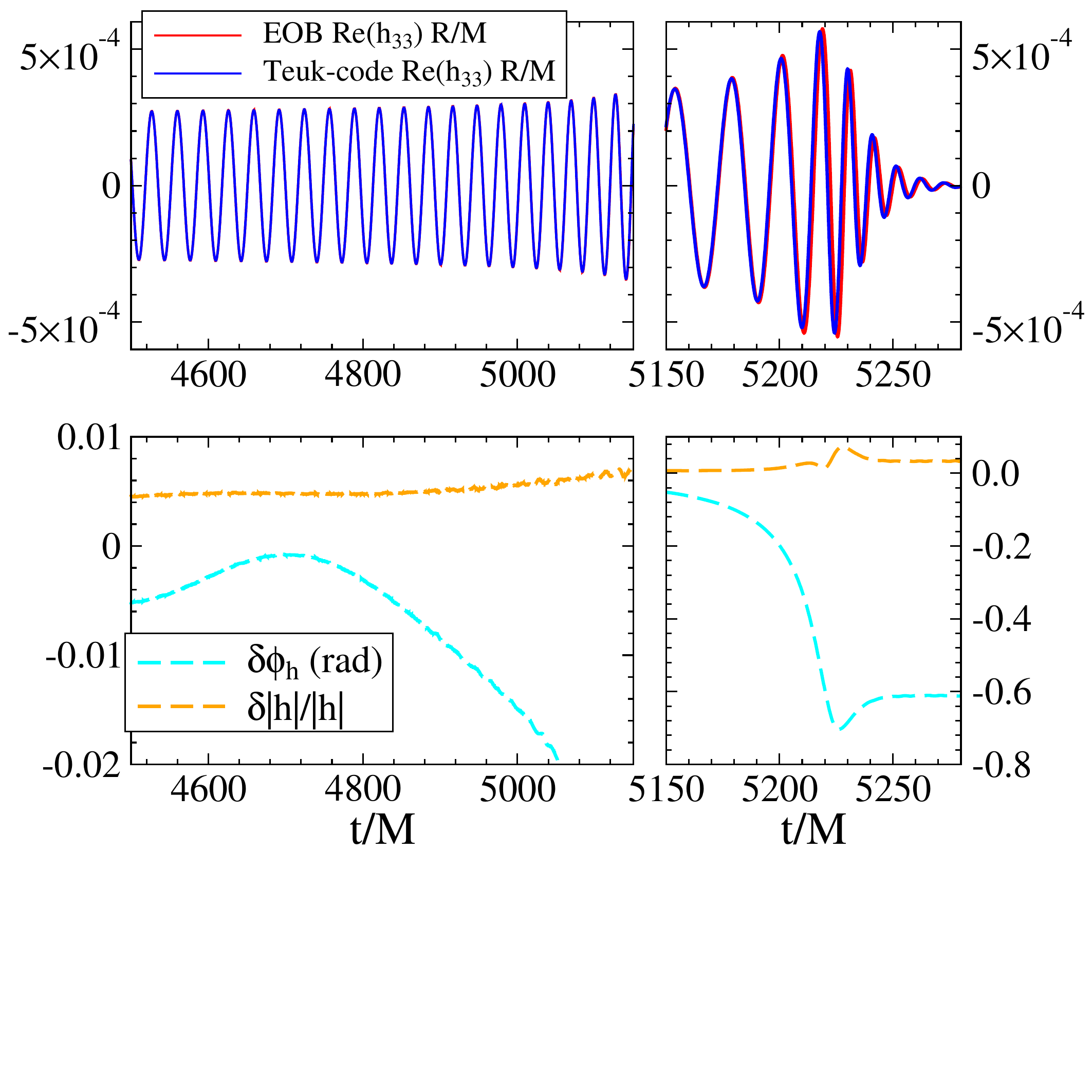} 
\vspace{-1.5cm}
\caption{\label{fig:modes22-33} Comparison between (2,2) and (3,3)
  modes generated by the Teukolsky code for $a=0$, and the
  corresponding modes produced with the EOB model of
  Ref.~\cite{Pan:2011gk}.  The EOB model was calibrated to
  numerical-relativity simulations at mass ratios $1, 1/2, 1/3, 1/4,
  1/6$, and extrapolated to $\nu = 10^{-3}$.  Upper panels show the
  real part of the modes; lower panels show the phase and fractional
  amplitude differences.}
\end{figure*}

\begin{enumerate}
\item The time at which the EOB $h_{22}$ reaches its peak should
  coincide with the time at which the EOB orbital frequency $\Omega$
  reaches its peak. We denote this time with
  $t_\mathrm{peak}^{\Omega}$.  The peaks of higher-order numerical
  modes differ from the peak of the numerical $h_{22}$; we define this
  time difference as
  \begin{eqnarray}\label{delta_orig}
    \Delta t_\mathrm{peak}^{\ell m} &=& t_\mathrm{peak}^{\ell m} -
    t_\mathrm{peak}^{22} \nonumber\\ &=& t_{d|h^\mathrm{NR}_{\ell
        m}|/dt=0} - t_{d|h^\mathrm{NR}_{22}|/dt=0}\,.
  \end{eqnarray}
  We require that the peaks of the EOB $h_{\ell m}$ occur at the time
  $t_\mathrm{peak}^{\Omega}+\Delta t_\mathrm{peak}^{\ell m}$:
  \begin{equation}
\left. \frac{d\left|h_{\ell m}^\mathrm{
    EOB}\right|}{dt}\right|_{t_\mathrm{peak}^{\Omega}+\Delta
  t_\mathrm{peak}^{\ell m}} =0\,.
  \end{equation}

 \item The peak of the EOB $h_{\ell m}$ should have the same amplitude
  as the peak of the numerical $h_{\ell m}$:
  \begin{equation}
    \left|h_{\ell m}^\mathrm{EOB}(t_\mathrm{peak}^{\Omega}+\Delta t_\mathrm{
        peak}^{\ell m})\right|=\left|h_{\ell m}^\mathrm{NR}(t_\mathrm{
        peak}^{\ell m})\right|.
  \end{equation}

 \item The peak of the EOB $h_{\ell m}$ should have the same second
   time derivative as the peak of the numerical $h_{\ell m}$:
  \begin{equation}
    \left. \frac{d^2\left|h_{\ell m}^\mathrm{
            EOB}\right|}{dt^2}\right|_{t_\mathrm{peak}^{\Omega}+\Delta
      t_\mathrm{peak}^{\ell m}} = \left. \frac{d^2\left|h_{\ell m}^\mathrm{
            NR}\right|}{dt^2} \right|_{t_\mathrm{peak}^{\ell m}}.
  \end{equation}

 \item The frequency of the numerical and EOB $h_{\ell m}$ waveforms
  should coincide at their peaks:
  \begin{equation}
    \omega^\mathrm{EOB}_{\ell m}(t_\mathrm{peak}^{\Omega}+\Delta t_\mathrm{
      peak}^{\ell m})=\omega^\mathrm{NR}_{\ell m}(t_\mathrm{peak}^{\ell m})\,.
  \end{equation}

 \item The time derivative of the frequency of the numerical and EOB
  $h_{\ell m}$ waveforms should coincide at their peaks:
  \begin{equation}
    \dot{\omega}^\mathrm{EOB}_{\ell m}(t_\mathrm{peak}^{\Omega}+\Delta
    t_\mathrm{peak}^{\ell m})=\dot{\omega}^\mathrm{NR}_{\ell m}(t_\mathrm{
      peak}^{\ell m})\,.
  \end{equation}

\end{enumerate}
[Note that the quantities $h_{\ell m}^\mathrm{EOB}$ referenced in the
  above equations are the same as the quantities $h^{\rm
    insp-plunge}_{\ell m}$ defined in Eq.~\eqref{hip}.]

The functions $\Delta t_\mathrm{peak}^{\ell m}$, $|h_{\ell
  m}^\mathrm{NR}(t_\mathrm{peak}^{\ell m})|$, $\left. d^2|h_{\ell
  m}^\mathrm{NR}|/dt^2 \right|_{t_\mathrm{ peak}^{\ell m}}$,
$\omega^\mathrm{NR}_{\ell m}(t_\mathrm{peak}^{\ell m})$, and
$\dot{\omega}^\mathrm{NR}_{\ell m}(t_\mathrm{peak}^{\ell m})$
described in Ref.\ {\cite{Pan:2011gk}} were extracted from
numerical-relativity and Teukolsky data, and approximated by smooth
functions of the symmetric mass ratio $\nu$.  Least-square fits for
these quantities were given in Table III of Ref.~\cite{Pan:2011gk}.
These fits included information about the $\nu = 10^{-3}$ case from
the analysis of this paper (which was in preparation as
Ref.\ {\cite{Pan:2011gk}} was completed).

\begin{figure*}
\includegraphics[width=8cm,clip=true]{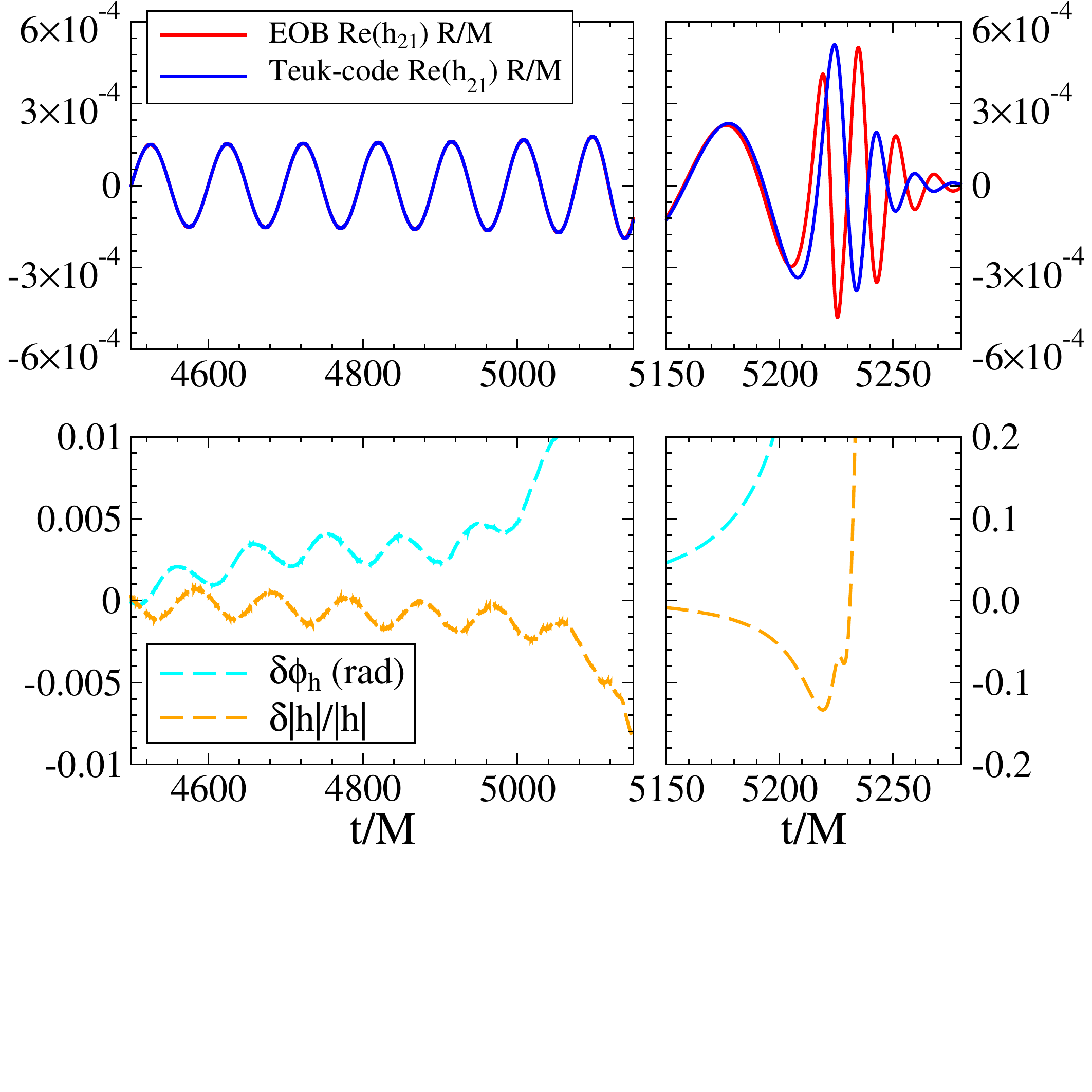} \qquad \qquad
\includegraphics[width=8cm,clip=true]{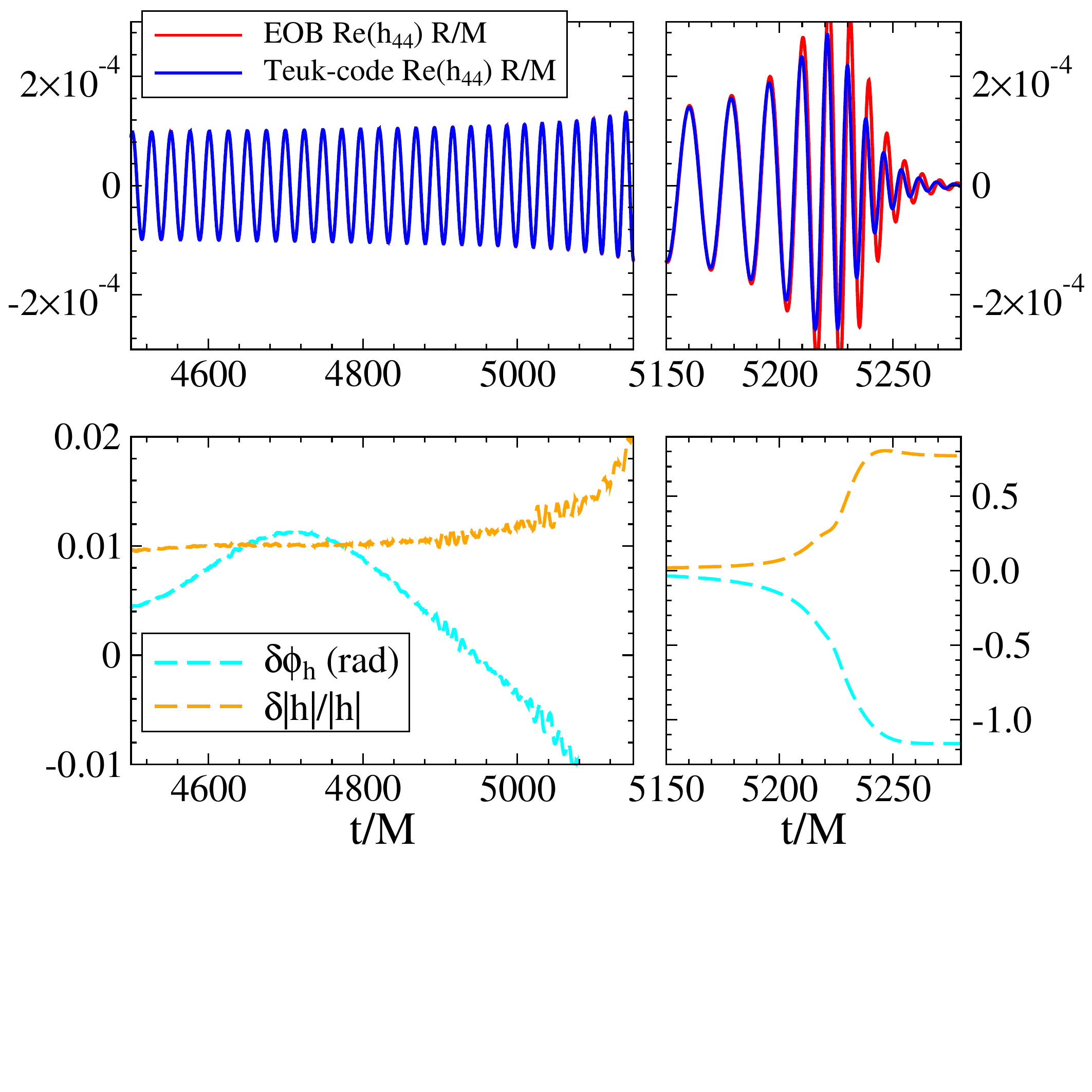} 
\vspace{-1.5cm}
\caption{\label{fig:modes21-44} The same as Fig.~\ref{fig:modes22-33},
  but for the (2,1) and (4,4) modes.}
\end{figure*}

Since $|h_{\ell m}(t_\mathrm{peak}^{\ell m})|$ and $\left. d^2|h_{\ell
  m}|/dt^2 \right|_{t_\mathrm{peak}^{\ell m}}$ approach zero in the
test-particle limit, their input values at $\nu=10^{-3}$ do not affect
the least-square fits very much. For the $(2,2)$ and $(3,3)$ modes,
the data points are very regular.  This is illustrated for the $(2,2)$
case in Fig.~\ref{fig:peakvaluefit}.  The residues of the fit at
$\nu=10^{-3}$ are very small, and the $\nu$-fits agree well with the
values from the Teukolsky code at $\nu=10^{-3}$.  Unfortunately, this
is not the case for the $(2,1)$ and $(4,4)$ modes.
Figure~\ref{fig:peakvaluefit} shows this for the $(4,4)$ mode.  The
$|h_{44}(t_\mathrm{peak}^{44})|$ data points do not lie on a smooth
curve, and so the fit is intrinsically unstable.  By minimizing the
relative residue instead of the absolute residue in the least-square
fit, we increase the weight on the $\nu=10^{-3}$ data point and get a
much better fit at low mass ratio, but at the cost of a much poorer
fit in the comparable-mass regime.  The situation is even worse for
$\omega_{44}(t_\mathrm{peak}^{44})$, for which the data points for
comparable masses have a rather irregular trend.  These results
emphasize the need for more accurate numerical-relativity data
describing the higher-order modes in order to smoothly connect these
quantities from the test-particle limit to the equal-mass case.

Once the coefficients $a^{h_{\ell m}}_{i}$ and $b^{h_{\ell m}}_{i}$
are known, we calculate $h^{\rm insp-plunge}_{\ell m}$ using
Eq.~\eqref{hip}, and attach the QNMs using Eq.~\eqref{eobfullwave}.
We assume the following comb widths~\cite{Pan:2011gk},
\begin{subequations}
  \label{combcal}
  \begin{eqnarray}
    &&\Delta t_\mathrm{match}^{22}= 5\, M \,, \qquad \!\! \Delta t_\mathrm{match}^{33}= 12\, M \,,\\
    && \Delta t_\mathrm{match}^{44}= 9\, M \,, \qquad\!\! \Delta t_\mathrm{match}^{21}= 8\, M \,,
  \end{eqnarray}
\end{subequations}
and choose $t_{\rm match}^{\ell m}=t_{\rm peak}^{\Omega}+\Delta t_{\rm
  peak}^{\ell m}$ with $\Delta t_{\rm peak}^{\ell m}$ given in
Eq.~(\ref{delta_orig}).  It was found in Ref.~\cite{Pan:2011gk} that,
after calibrating the EOB adjustable parameters and aligning the EOB
and numerical waveforms at low frequency, the difference between
$t_{\rm peak}^{22}$ and $t_{\rm peak}^{\Omega}$ is typically $\sim
1M$. Thus, Ref.~\cite{Pan:2011gk} assumed that $t_{\rm
  peak}^{\Omega}=t_{\rm peak}^{22}$ and consequently set $\Delta
t^{22}_{\rm peak}=0$.  Following these findings, we attach the QNMs at
$t_{\rm peak}^{\Omega}$ for the $(2,2)$ mode, and at $t_{\rm
  peak}^{\Omega}+\Delta t_{\rm peak}^{\ell m}$ for all the other
modes.

Figures {\ref{fig:modes22-33}} and {\ref{fig:modes21-44}} compare the
leading modes generated by this EOB model with the modes generated by
the time-domain Teukolsky code.  We adopt the waveform alignment
procedure used in Refs.~\cite{Boyle2008a, Buonanno:2009qa,
  Pan2010hz,Pan:2011gk} and Sec.~\ref{sec:FD-TD}, aligning the
waveforms at low frequency by minimizing
\begin{equation}\label{waveshifts}
\int_{t_1}^{t_2}\left[\phi_1(t)-
    \phi_2(t-\Delta t)-\Delta\phi\right]^2\,dt\,,
\end{equation}
over a time shift $\Delta t$ and a phase shift $\Delta\phi$.  Here,
$\phi_1(t)$ and $\phi_2(t)$ are the gravitational phases of the EOB
and Teukolsky $h_{22}$.  We chose $t_2 - t_1 = 1000M$, and center these
times when the orbital frequencies are low. We have verified that our results
are insensitive to the precise location of this integration interval, provided that it is
chosen during the inspiral phase.

As expected from the discussion above, Figure {\ref{fig:modes22-33}}
shows that there is quite good agreement between the EOB and Teukolsky
models for the $(2,2)$ and $(3,3)$ modes.  In particular, the
difference in both the amplitude and the phase is quite small until
the inspiral reaches the ISCO. This excellent agreement is due to the
resummed-factorized energy flux~\cite{DIN} employed in the EOB
equations of motion and waveforms (see previous
studies~\cite{Nagar:2006xv,Damour2007,Pan:2010hz,Fujita:2010xj,
  Bernuzzi:2010ty,Bernuzzi:2010xj,Fujita:2011zk,Bernuzzi:2011aj}).
The amplitude disagreement during merger and ringdown is due to our
procedure of attaching QNMs to the EOB waveform [see Fig.~3 and
  discussion around in Ref.~\cite{Pan:2011gk}].  The accumulation of
some phase difference during plunge will be discussed at the end
of this section.

\begin{figure*}
\includegraphics[width=8cm,clip=true]{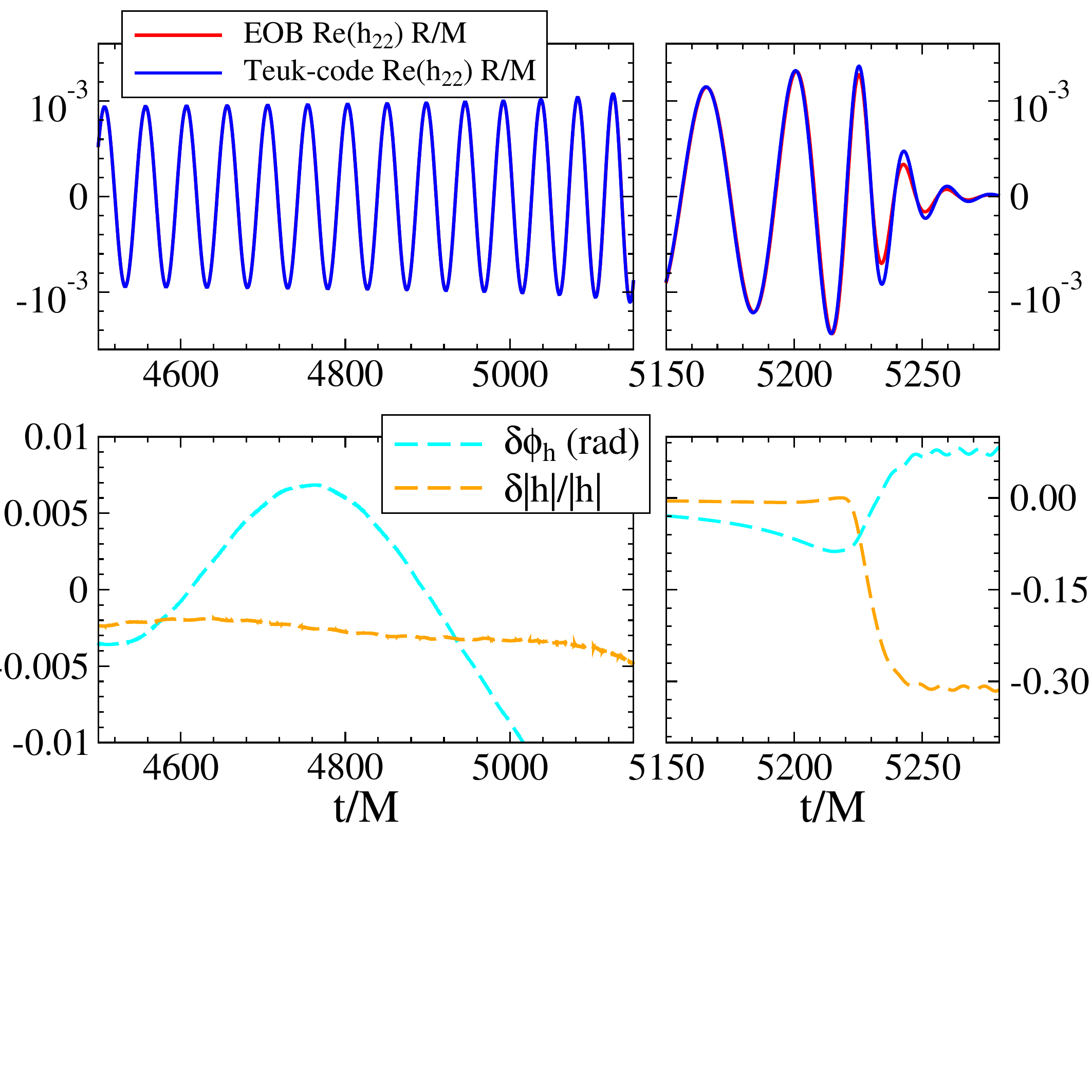} \qquad
\includegraphics[width=8cm,clip=true]{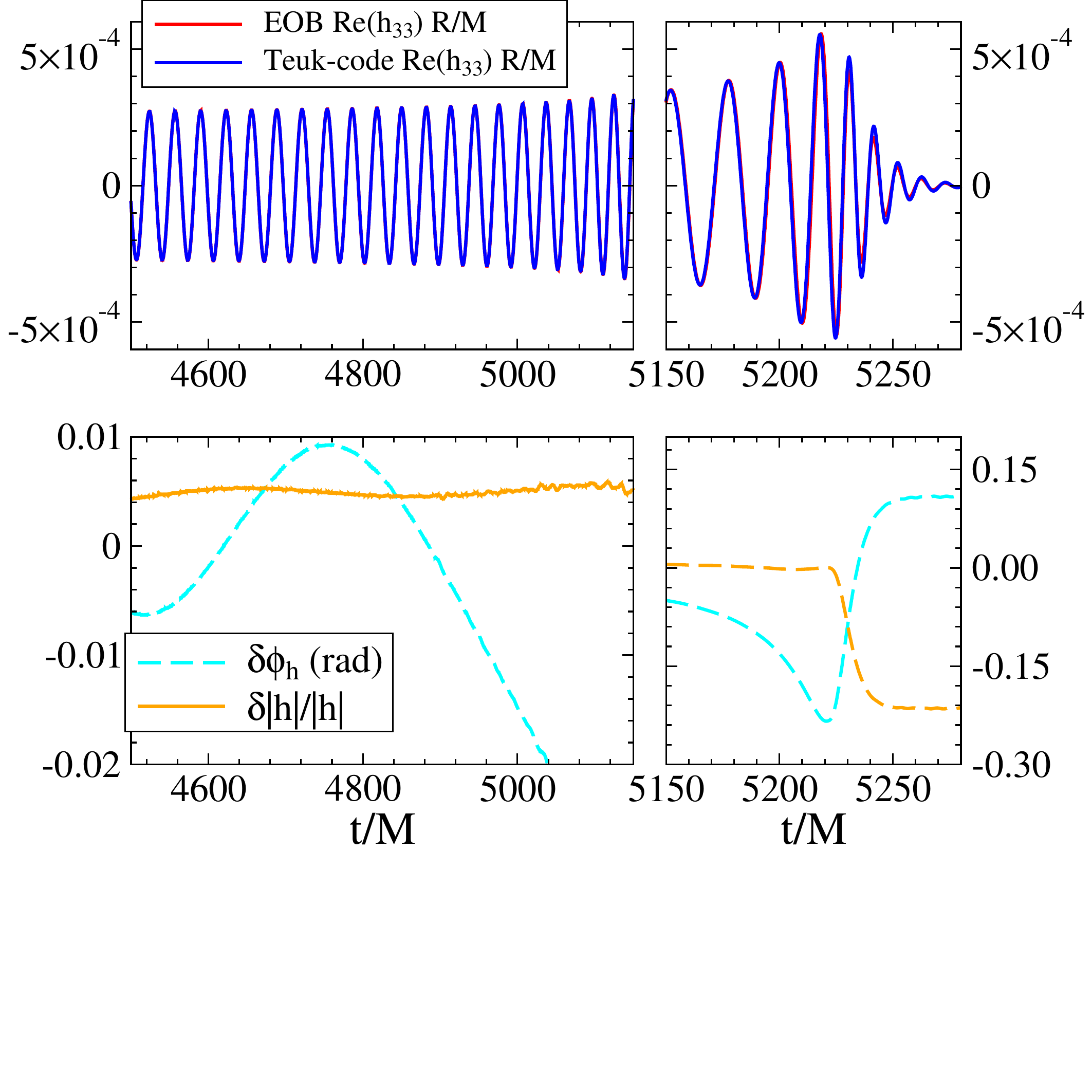} 
\vspace{-1.5cm}
\caption{\label{fig:modes22-33Cal} The same as
  Fig.\ {\ref{fig:modes22-33}}, but for the Teukolsky-calibrated EOB
  model described in Sec.\ \ref{sec:results_improved}.}
\end{figure*}

Figure \ref{fig:modes21-44} shows that the agreement between the EOB
and Teukolsky $(2,1)$ and $(4,4)$ modes remains excellent during the
long inspiral, but is not very satisfactory during the merger and
ringdown.  For the $(4,4)$ mode, the EOB amplitude becomes too large
toward merger.  This is a consequence of the excessively large residue
of the $\nu$-fit for $|h_{44}(t_\mathrm{peak}^{44})|$ at $\nu =
10^{-3}$.  For the $(2,1)$ mode, the EOB model of
Ref.~\cite{Pan:2011gk} fails to reproduce a reasonable merger
waveform.  This problem is related to the fact that the value of
$\Delta t_{\rm peak}^{21}$ at $\nu = 10^{-3}$ used in
Ref.\ {\cite{Pan:2011gk}} to determine the $\nu$-fit is too large (see
Table~\ref{tab:delay_nonspinning}).  The problem is deeper than this,
however.  In particular, the value is uncertain due to the (unusual)
broadness of the Teukolsky $(2,1)$ mode's peak (see
Fig.~\ref{fig:modesampl}).  We shall see in the next section that to
improve the agreement of the $(2,1)$ mode, we need a smaller value for
$\Delta t_{\rm peak}^{21}$.

An additional source of error arises from the procedure that was used
to compute the NQC coefficients $a_1^{h_{22}}$, $a_2^{h_{22}}$, and
$a_3^{h_{22}}$ used in $N_{22}$ [see Eq.~(\ref{Nlm})].  In
Ref.\ {\cite{Pan:2011gk}}, these coefficients were calculated by an
iterative procedure using the five conditions discussed at the
beginning of this section.  These coefficients have small but
non-negligible effects on the EOB dynamics: through the amplitude
$|h_{22}|$, they enter the energy flux [see Eq.~\eqref{resflux}] and
thereby influence the rate at which the small body spirals
in\footnote{Note that the NQC coefficients $a_i^{h_{\ell m}}$ of
  higher-order modes contribute much less to the energy flux and can
  be safely ignored in the dynamics of comparable-mass binaries
  \cite{Pan:2011gk}.}.  This iterative procedure increases by a factor
of a few the computational cost of generating $h_{22}$.  To mitigate
this cost increase, Ref.~\cite{Pan:2011gk} suggested replacing the
iterative procedure with $\nu$-fits for $a_1^{h_{22}}$, $a_2^{h_{22}}$
and $a_3^{h_{22}}$.  These fits were obtained using data for mass
ratios $1,1/2,1/3,1/4$ and $1/6$.  The EOB waveforms shown in
Figs.~\ref{fig:modes22-33}, \ref{fig:modes21-44} are then generated
using these $\nu$-fits extrapolated to $\nu = 10^{-3}$.  When
comparing the fit and the true values of $a_1^{h_{22}}$,
$a_2^{h_{22}}$ and $a_3^{h_{22}}$ at $\nu = 10^{-3}$, we find a
non-negligible difference which is responsible for $\sim 0.4$ rad
difference between the EOB and Teukolsky $(2,2)$ modes, and for $\sim
0.6$ rad difference for the $(3,3)$ modes.  In the next section, we
shall show that by returning to the iterative procedure, rather than
using the fits, we can do much better.

Lastly, we comment on why for the $(2,1)$ and $(4,4)$ modes in
Eq.~\eqref{change} we replaced $v_\Phi^{(\ell+\epsilon)}$ with
$v_\Phi^{(\ell+\epsilon-2)}/r_\Omega$.  As discussed above, the
amplitude of the numerical $(2,1)$ and $(4,4)$ modes reaches a peak a
fairly long time after the peak of the $(2,2)$ mode. Thus, in order to
impose the first condition (in our list of five) given above, the peak
of the EOB mode should be moved to $t_\mathrm{peak}^{\Omega}+\Delta
t_\mathrm{peak}^{\ell m}$. However, the leading EOB amplitude is
proportional to a power of the orbital frequency.  This frequency
decreases to zero at the horizon, and so the EOB amplitude drops to an
extremely small value at $t_\mathrm{peak}^{\Omega}+\Delta
t_\mathrm{peak}^{\ell m}$.  By replacing $v_\Phi^2=(M r_\Omega  \Omega)^2$
with $1/r_\Omega$, we slow the decay of these modes after
$t_\mathrm{peak}^{\Omega}$, and can successfully move the peak of the
mode to $t_\mathrm{ peak}^{\Omega}+\Delta t_\mathrm{peak}^{\ell m}$.
This modification was also adopted in Ref.~\cite{Pan:2011gk} to
successfully model the $(2,1)$ and $(4,4)$ modes in the
comparable-mass case.

\subsection{Comparison of calibrated EOB waveforms and Teukolsky waveforms for $a = 0$}
\label{sec:results_improved}

\begin{figure*}
\includegraphics[width=8cm,clip=true]{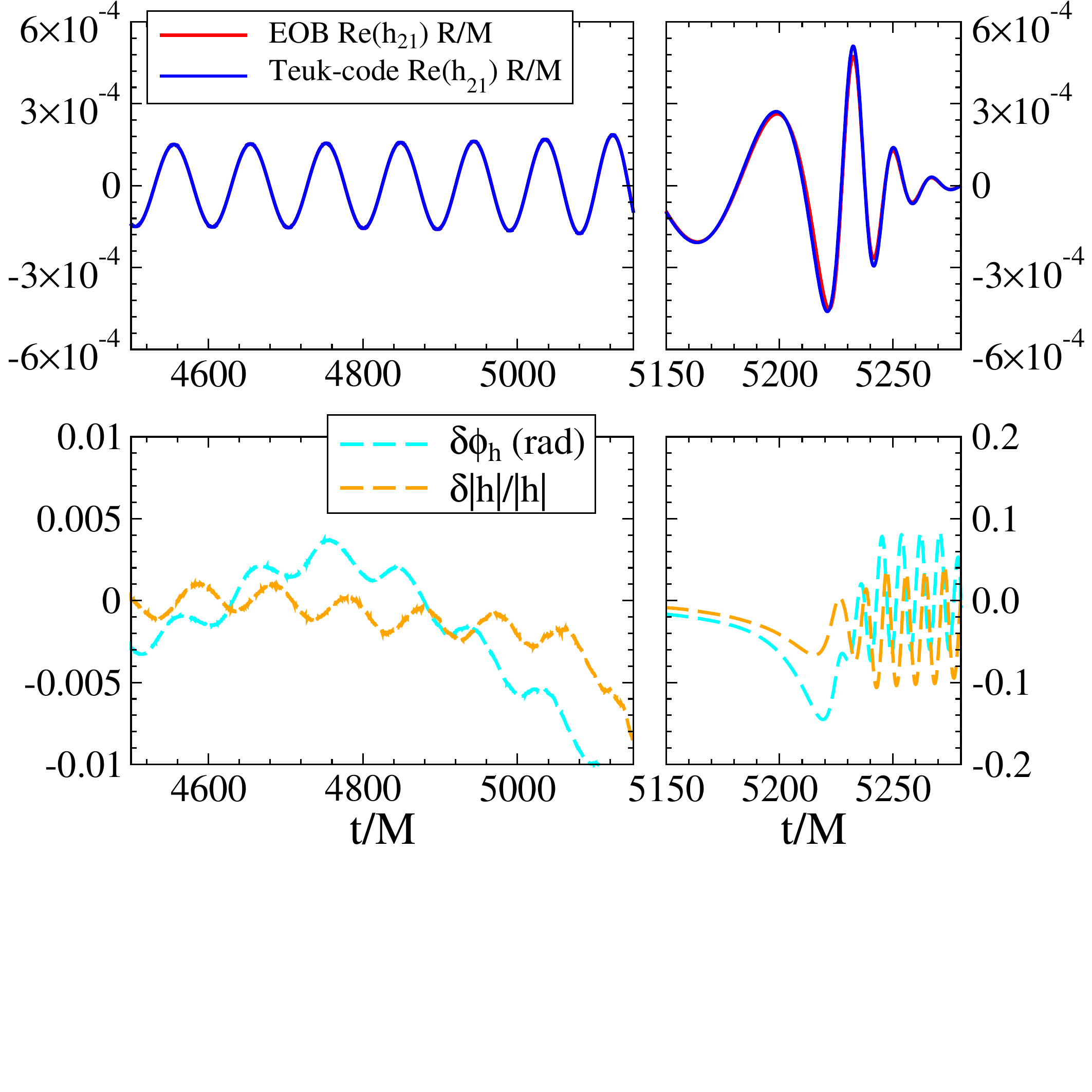} \qquad 
\includegraphics[width=8cm,clip=true]{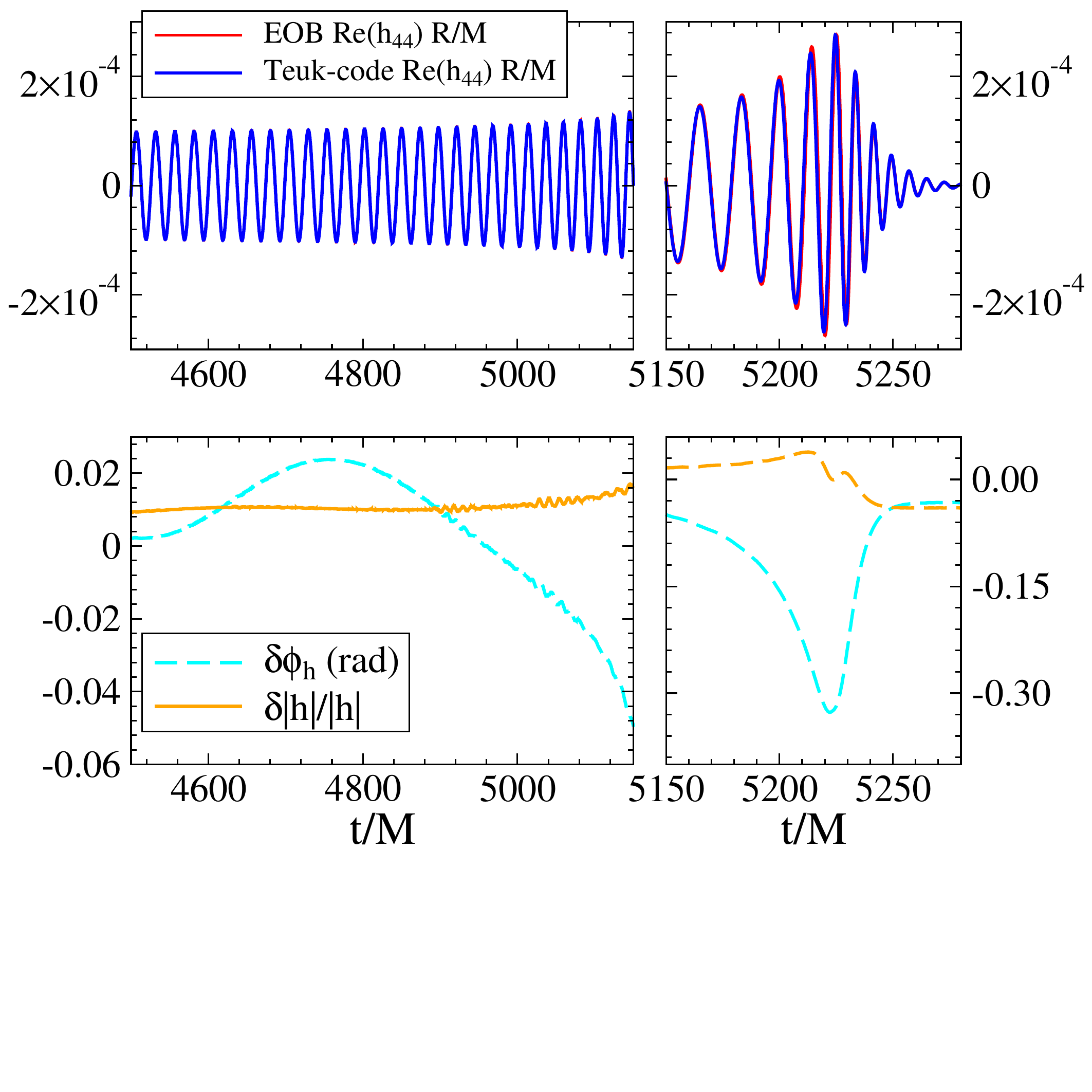} 
\vspace{-1.5cm}
\caption{\label{fig:modes21-44Cal} The same as
  Fig.~\ref{fig:modes21-44}, but for the Teukolsky-calibrated EOB
  model described in Sec.~\ref{sec:results_improved}.}
\end{figure*}

We now improve on the EOB model of Ref.\ {\cite{Pan:2011gk}} to more
accurately reproduce Teukolsky waveforms.  We focus on comparisons for
the $a = 0$ limit, although we discuss how we build our model for
general spins.  We start from the five conditions discussed at the
beginning of Sec.~\ref{sec:results_original} which allow us to compute
the NQC coefficients:
\begin{eqnarray}
\label{cond1}
\left. \frac{d\left|h_{\ell m}^\mathrm{
            EOB}\right|}{dt}\right|_{t_\mathrm{peak}^{\Omega}+\Delta
      t_\mathrm{peak}^{\ell m}} &=&0\,\\
\label{cond2} \left|h_{\ell m}^\mathrm{EOB}(t_\mathrm{peak}^{\Omega}+\Delta t_\mathrm{
        peak}^{\ell m})\right| &=& \left|h_{\ell m}^\mathrm{Teuk}(t_\mathrm{
        peak}^{\ell m})\right|\,,\\
\label{cond3} \left. \frac{d^2\left|h_{\ell m}^\mathrm{
            EOB}\right|}{dt^2}\right|_{t_\mathrm{peak}^{\Omega}+\Delta
t_\mathrm{peak}^{\ell m}} &=& \left. \frac{d^2\left|h_{\ell m}^\mathrm{
            Teuk}\right|}{dt^2} \right|_{t_\mathrm{peak}^{\ell m}}\,,\\
\label{cond4} \omega^\mathrm{EOB}_{\ell m}(t_\mathrm{peak}^{\Omega}+\Delta t_\mathrm{
      peak}^{\ell m}) &=& \omega^\mathrm{Teuk}_{\ell m}(t_\mathrm{peak}^{\ell m})\,,\\
\label{cond5} \dot{\omega}^\mathrm{EOB}_{\ell m}(t_\mathrm{peak}^{\Omega}+\Delta
    t_\mathrm{peak}^{\ell m}) &=& \dot{\omega}^\mathrm{Teuk}_{\ell m}(t_\mathrm{
      peak}^{\ell m})\,,
\end{eqnarray}
where
\begin{eqnarray}
\label{delta_new}
\Delta t_\mathrm{peak}^{\ell m} &=& t_\mathrm{peak}^{\ell m} - t_\mathrm{peak}^{\Omega}\,\nonumber\\
    &=& t_{d|h^\mathrm{Teuk}_{\ell m}|/dt=0} - t_{d\Omega/dt=0}\,.
  \end{eqnarray}
As before, the quantities $h_{\ell m}^\mathrm{EOB}$ given above are
equivalent to the quantities $h^{\rm insp-plunge}_{\ell m}$ in
Eq.~\eqref{hip}.  The quantities on the RHSs of
Eqs.~(\ref{cond1})--(\ref{cond5}) are now computed from the Teukolsky
waveforms, rather than using the $\nu$-fits of Ref.~\cite{Pan:2011gk},
as in Sec.~\ref{sec:results_original}.  Notice that $\Delta
t_\mathrm{peak}^{\ell m}$ in Eq.~(\ref{delta_new}) differs from the
one in Eq.~(\ref{delta_orig}). In fact, since the Teukolsky code uses
the EOB trajectory, the time difference between the peak of any
$(\ell,m)$ mode and the peak of the orbital frequency is unambiguous.
This was not the case in Ref.~\cite{Pan:2011gk} where numerical
relativity waveforms are used.  In that case, the time difference
between the peak of the numerical relativity $(\ell,m)$ modes and the
peak of the EOB orbital frequency depends on the alignment procedure
between the numerical and EOB waveforms, and on the calibration of the
EOB adjustable parameters.  This is why Ref.~\cite{Pan:2011gk} adopted
the $\Delta t_\mathrm{peak}^{\ell m}$ described in
Eq.~(\ref{delta_orig}).

As discussed in Sec.~\ref{sec:results_original},
Ref.~\cite{Pan:2011gk} assumed that $t_{\rm peak}^{\Omega}=t_{\rm
  peak}^{22}$ and consequently set $\Delta t^{22}_{\rm peak}=0$.
However, as seen in Tables~\ref{tab:delay_nonspinning},
\ref{tab:delay_spinning}, the Teukolsky data show that $t_{\rm
  peak}^{\Omega}-t_{\rm peak}^{22}$ differs from zero when $\nu =
10^{-3}$; this effect is particularly pronounced for large positive
Kerr spin parameters.  The modified prescription given by
Eq.~(\ref{delta_new}) is thus quite natural.  By contrast, the
prescription described in Ref.~\cite{Pan:2011gk} is bound to fail for
all the modes in the test-particle limit.
We point out that in Appendix A of Ref.~\cite{Bernuzzi:2010xj}  the authors explored how
non-spinning waveforms can be improved,  by attaching the QNMs at a shifted
time (common to all modes) $\sim 3M$ past the peak of the orbital frequency, while maintaining
 the determination of the NQC parameters at the peak of the orbital
 frequency.

In the non-spinning case, we solve Eqs.~\eqref{cond1}, \eqref{cond2}
for $a^{h_{\ell m}}_{i}$ with $i=1,2,3$, and set $a^{h_{\ell
    m}}_{4}=a^{h_{\ell m}}_{5}=0$. In the spinning case, in order to
not introduce spin-dependence at leading order, we fix $a^{h_{\ell
    m}}_{1}$ and $a^{h_{\ell m}}_{2}$ to the values calculated for $a
= 0$, and solve for $a^{h_{\ell m}}_{i}$ (with $i=3,4,5$).
As for the phase NQC coefficients, in the non-spinning case we solve
Eqs.\ \eqref{cond3} and \eqref{cond4} for $b^{h_{\ell m}}_{i}$ (with $i=1,2$) and set $b^{h_{\ell m}}_{3} =
b^{h_{\ell m}}_4 = 0$.  In the spinning case, we fix 
$b^{h_{\ell m}}_{1}$ and $b^{h_{\ell m}}_{2}$ to their $a = 0$ values (again, in order to not introduce
any spin-dependence at leading order),
and solve for $b^{h_{\ell m}}_{3}$ and $b^{h_{\ell m}}_{4}$.

Once the coefficients $a^{h_{\ell m}}_{i}$ and $b^{h_{\ell m}}_{i}$
are known, we calculate $h^{\rm insp-plunge}_{\ell m}$ using
Eq.~\eqref{hip} and attach the QNMs using Eq.~\eqref{eobfullwave},
assuming the comb's width as in Eq.~(\ref{combcal}).  Furthermore, we
choose $t_{\rm match}^{\ell m}=t_{\rm peak}^{\Omega}+\Delta t_{\rm
  peak}^{\ell m}$ with $\Delta t_{\rm peak}^{\ell m}$ given in
Eq.~(\ref{delta_new}).  The merger-ringdown $(2,2)$ mode is now
attached at the time where the Teukolsky $(2,2)$ amplitude peaks, in contrast
to the approach used in Ref.~\cite{Pan:2011gk}.  All the other
merger-ringdown $(\ell,m)$ modes are attached at the time the
corresponding Teukolsky $(\ell,m)$ amplitudes peaks, which is the same
procedure followed in Ref.~\cite{Pan:2011gk}.

\begin{table}
\centering 
\begin{tabular}{c| c| c| c| c| c} 
\hline\hline 
$(\ell,m)$ & $\Delta t^{\ell m}_{\rm peak}$ & $|{h}_{\ell m,\rm peak}^{\rm Teuk}|$ & $d^2{|h_{\ell m,\rm peak}^{\rm Teuk}|}/dt^2$ & $\omega_{\ell m,\rm peak}^{\rm Teuk}$ & 
$\dot{\omega}_{\ell m,\rm peak}^{\rm Teuk}$\T \B\\ 
\hline 
 (2,2) & -2.99 & 0.001450 & -3.171$\times10^{-6}$ & 0.2732 &
   0.005831 \\
 (2,1) & 6.32 & 0.0005199 & -7.622$\times10^{-7}$ & 0.2756 &
   0.01096 \\
 (3,3) & 0.52 & 0.0005662 & -1.983$\times10^{-6}$ & 0.4546 &
   0.01092 \\
 (4,4) & 2.26 & 0.0002767 & -1.213$\times10^{-6}$ & 0.6347 &
   0.01547 \\
\hline\hline
\end{tabular}
\caption{\label{tab:input} Input values for the RHSs of
  Eqs.~(\ref{cond1})--(\ref{cond5}) for the EOB model for non-spinning
  black-holes used in Figs.~\ref{fig:modes22-33Cal}
  and~\ref{fig:modes21-44Cal}.}
\end{table}

Figures \ref{fig:modes22-33Cal} and \ref{fig:modes21-44Cal} compare
these calibrated EOB models to the Teukolsky amplitudes.  Table
\ref{tab:input} lists the input parameters used on the RHSs of
Eqs.~(\ref{cond1})--(\ref{cond5}).  We emphasize that the value of
$\Delta t^{21}_{\rm peak}$ we reported in this table and that we use
in our model is $2.5 M$ smaller than the difference $t_{\rm
  peak}^{21}-t_{\rm peak}^{\rm \Omega}$.  We do this because the peak
of the Teukolsky $(2,1)$ amplitude is quite broad, and at the time
$t_{\rm peak}^{21}$ where the peak occurs the Teukolsky mode's
frequency oscillates due to superposition of the $\ell=2$, $m\pm1$
modes, as discussed in Sec.\ \ref{sec:characteristics}.  (See also
Ref.~\cite{Bernuzzi:2010ty} and Fig.\ 3 of
Ref.\ Ref.~\cite{Bernuzzi:2010xj}, which shows similar oscillations.)
Although these oscillations are physical, we do not attempt to
reproduce them in our EOB waveform, as their effect on the phase
agreement between the EOB and Teukolsky waveforms is negligible.  We
therefore simply choose a slightly smaller value of $\Delta
t^{21}_{\rm peak}$, ensuring that $t_{\rm peak}^{\Omega}+\Delta t_{\rm
  peak}^{21}$ is within the broad peak of $h_{\rm peak}^{21}$.  This
in turn ensures that the oscillations in frequency do not impact our
results.

\begin{figure*}
 \includegraphics[width=8cm,clip=true]{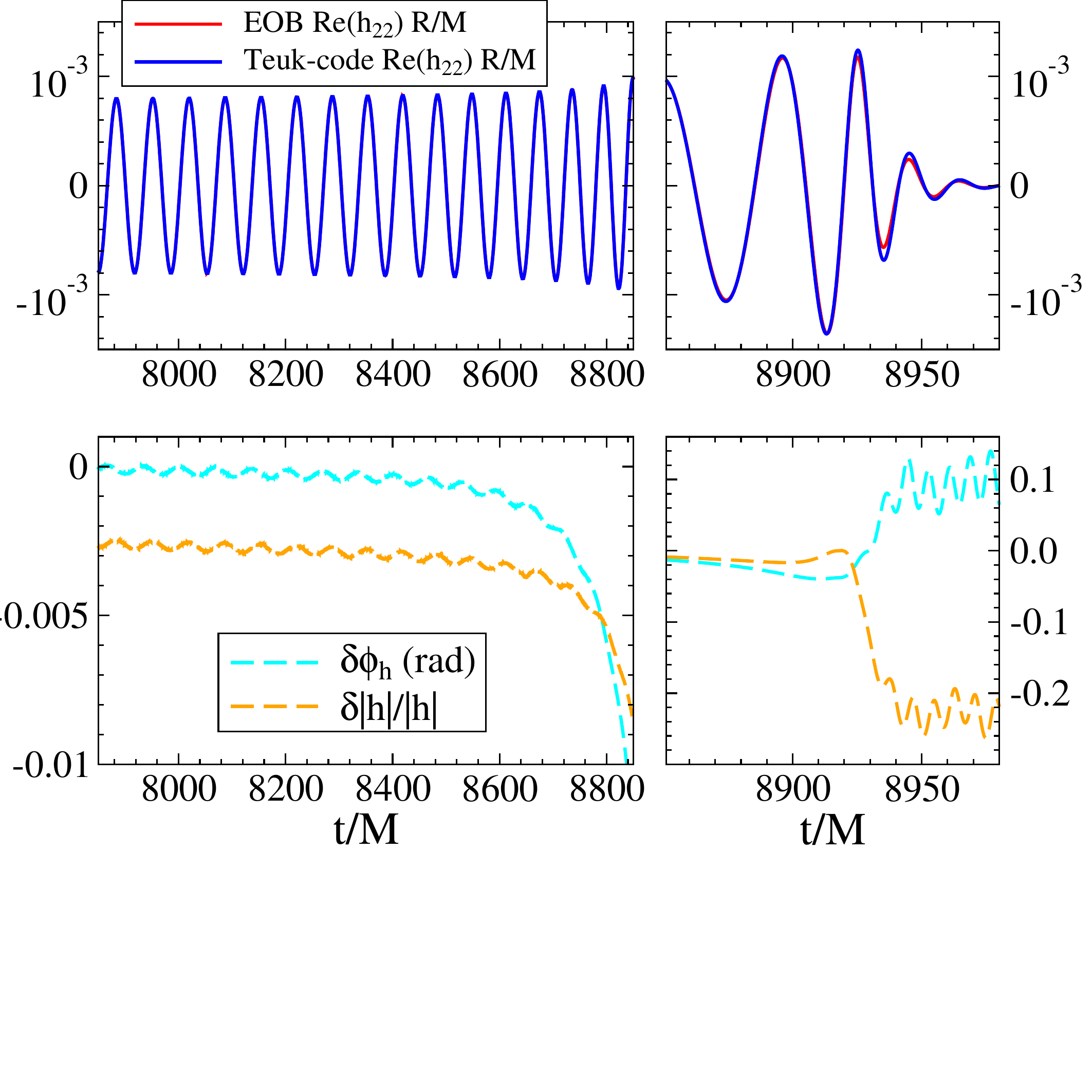} \qquad 
 \includegraphics[width=8cm,clip=true]{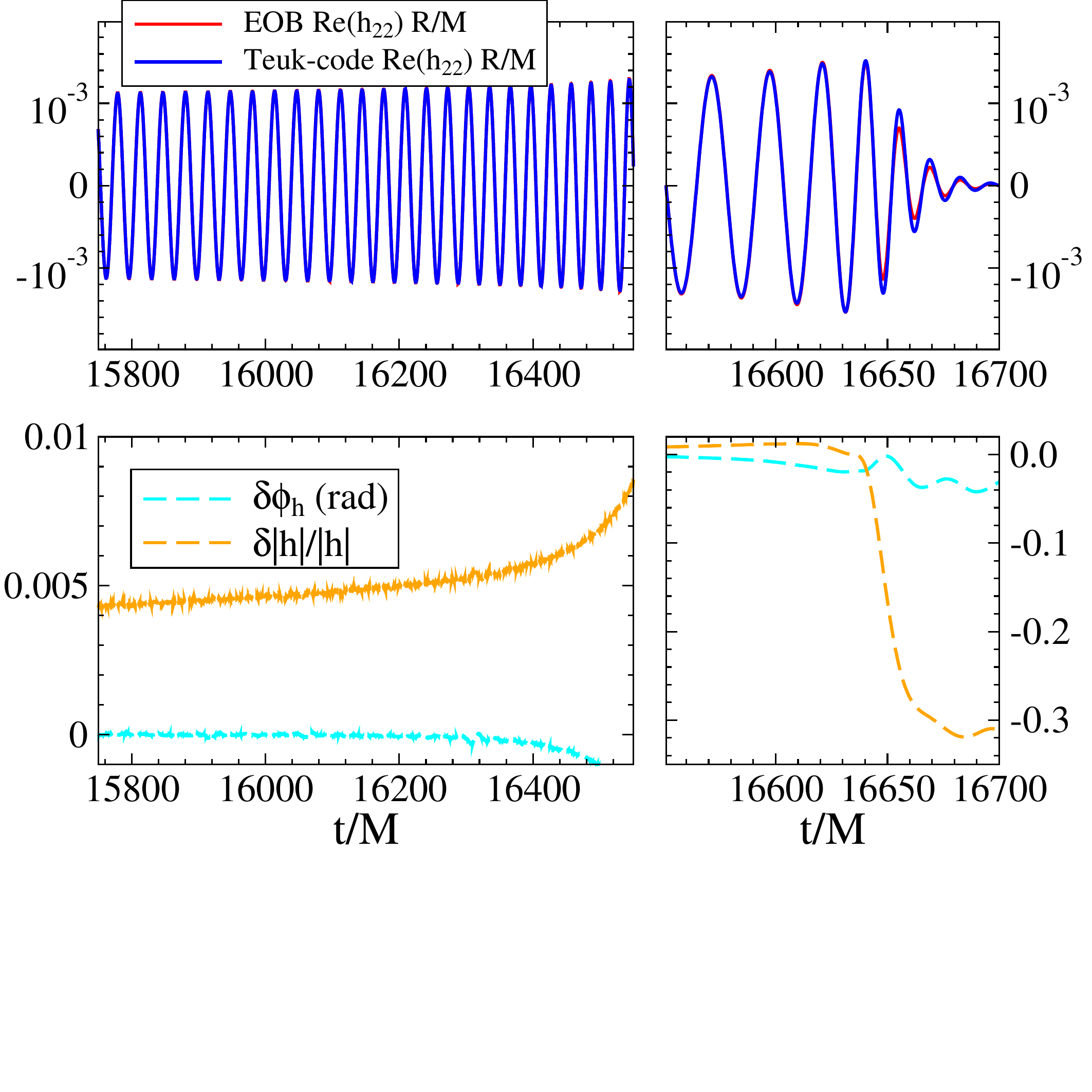} 
\vspace{-1.5cm}
\caption{\label{h22-Cal-Comparisons-a05} Comparison of
  Teukolsky-calibrated EOB and Teukolsky $(2,2)$ modes for $a/M=-0.5$
  (left panel) or $a/M=0.5$ (right panel).  Upper panels show the real
  part of the modes; lower panels show phase and fractional amplitude
  differences.}
\end{figure*}

Figure \ref{fig:modes21-44Cal} demonstrates that, by calibrating
against the Teukolsky waveforms using the input values shown in Table
\ref{tab:input}, the agreement between the EOB and Teukolsky waveforms
is considerably improved during merger and ringdown for the $(2,1)$
and $(4,4)$ modes.  Improvements to the $(2,2)$ and $(3,3)$ modes
(Fig.\ \ref{fig:modes22-33Cal}) are less significant, since the model
of Ref.~\cite{Pan:2011gk} works quite well for these modes.  As we
discussed in the previous section, the input parameters listed in 
Table~\ref{tab:input} are well predicted by the fitting formulas of
Ref.~\cite{Pan:2011gk} (see also Fig.~\ref{fig:peakvaluefit}).  There
is, however, noticeable improvement in phase agreement between
Figs.~\ref{fig:modes22-33} and \ref{fig:modes22-33Cal}.  This is due
to the fact that in the latter case we use the iterative procedure to
compute the NQC coefficients $a_i$, as discussed at the end of
Sec.~\ref{sec:results_original}.

Comparing to the discussion of numerical error in
Sec.~\ref{sec:num-err}, we note that the differences in phase and
amplitude between the EOB and Teukolsky modes shown in
Figs.~\ref{fig:modes22-33Cal} and \ref{fig:modes21-44Cal} are within
the numerical errors essentially through the plunge; the differences
grow larger than these errors during merger and ringdown.  As this
analysis was being completed, we acquired the capability to produce
Teukolsky waveforms using the hyperboloidal layer method
(Ref.\ {\cite{ZenKha2011}}; see also Ref.\ {\cite{Bernuzzi:2011aj}}).
We have compared phase and amplitude differences for the $(2,2)$ mode
between the Teukolsky code used for the bulk of this analysis, and the
hyperboloidal variant.  We have found that these differences are
within the errors discussed in Sec.~\ref{sec:num-err}.

\subsection{Comparisons for general spin}
\label{sec:results_kerr_eob}

\begin{figure*}
 \includegraphics[width=8cm,clip=true]{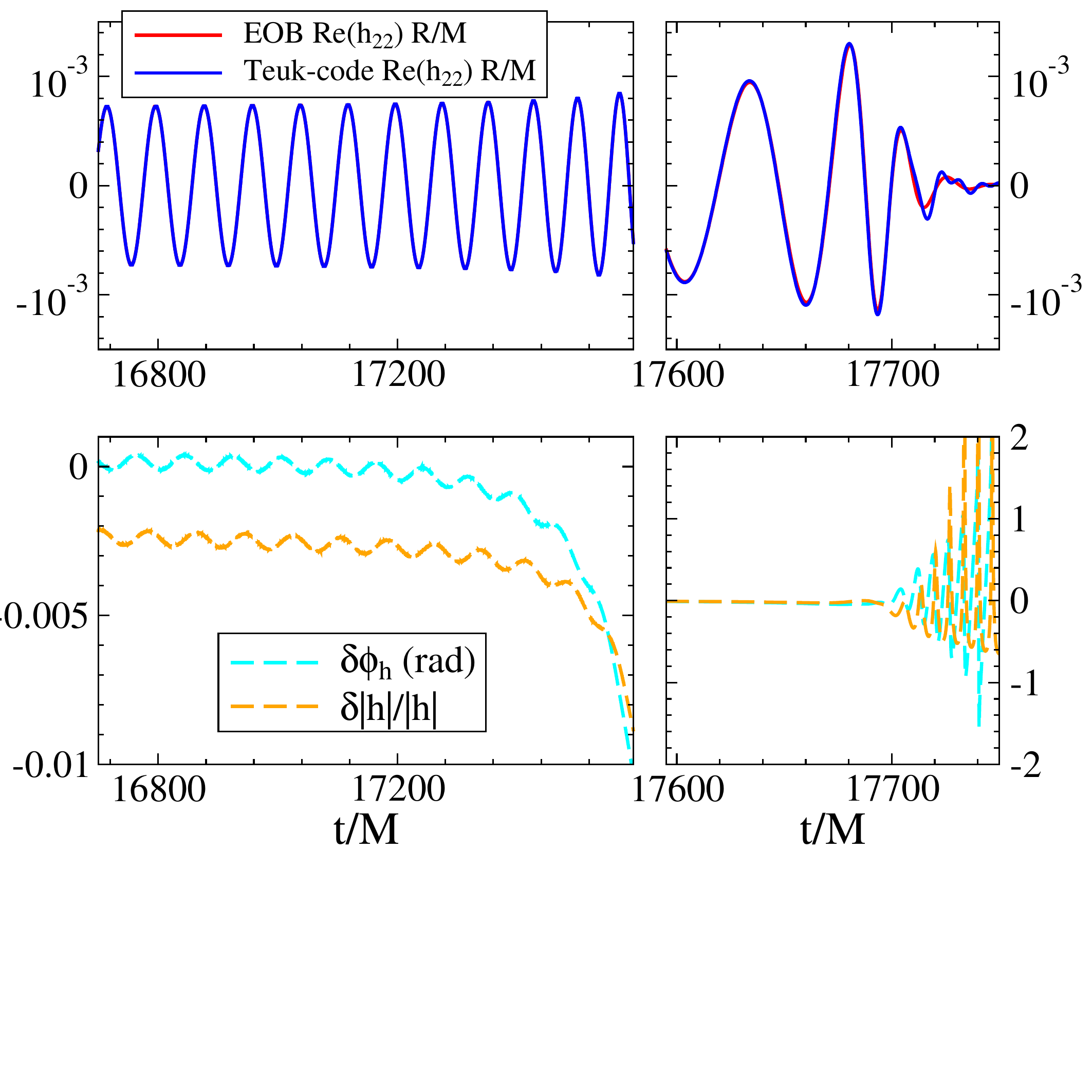} \qquad 
 \includegraphics[width=8cm,clip=true]{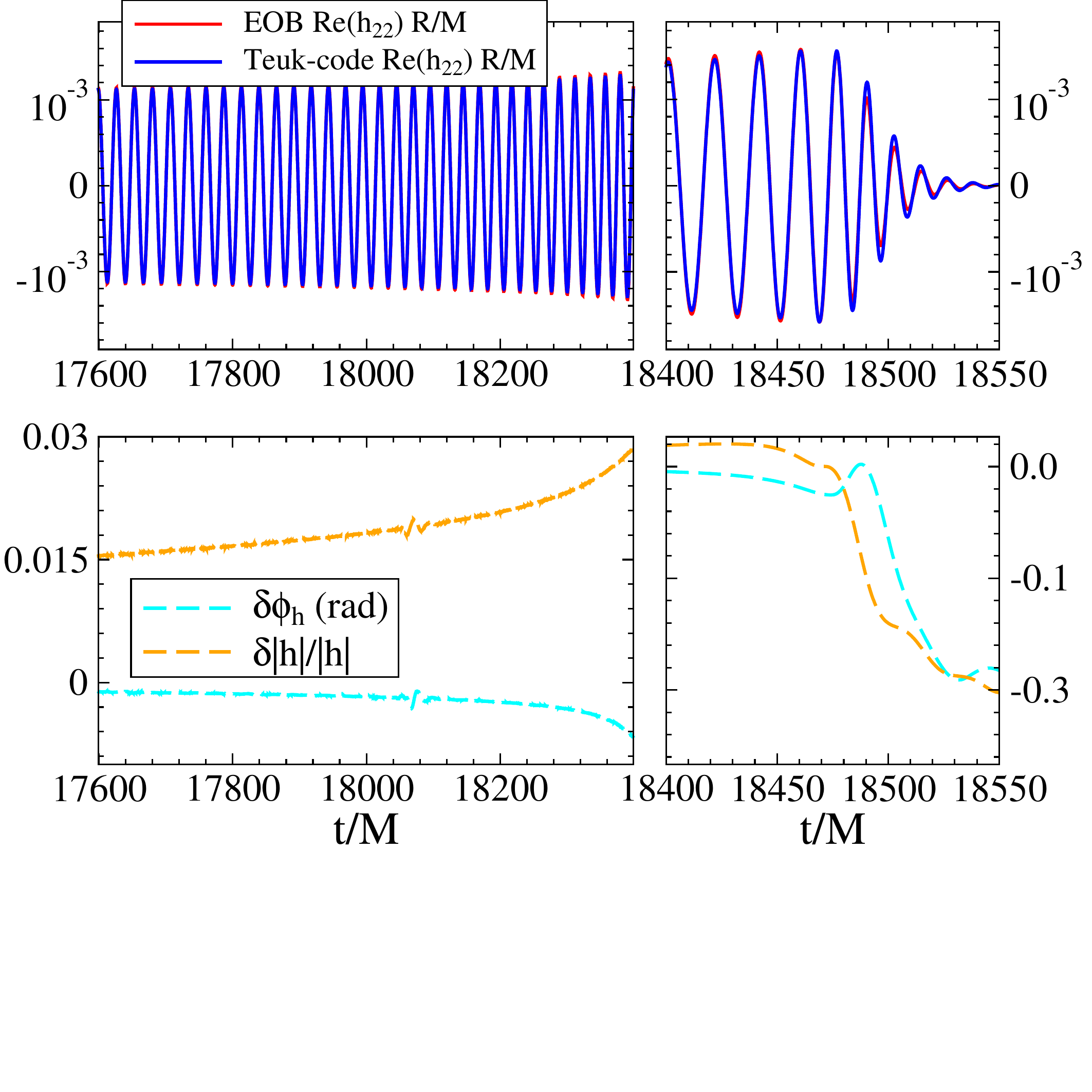} 
\vspace{-1.5cm}
\caption{\label{h22-Cal-Comparisons-a07} The same as
  Fig.~\ref{h22-Cal-Comparisons-a05}, but for $a/M=-0.9$ (left panel)
  and $a/M=0.7$ (right panel).}
\end{figure*}

We conclude our discussion of results by comparing, for the first
time, EOB and Teukolsky coalescence waveforms with $a\neq 0$.  These
waveforms are produced using the trajectory of the spinning EOB model
described in Sec.~\ref{sec:analytics}.  As in the non-spinning case,
understanding the transition from inspiral to ringdown in the
test-particle limit when the central black hole carries spin can help
modeling the plunge-merger waveforms from comparable-mass spinning
black holes. 

As in the non-spinning case~\cite{Nagar:2006xv,
  Damour2007, Pan:2010hz, Fujita:2010xj, Bernuzzi:2010ty,
  Bernuzzi:2010xj, Fujita:2011zk, Bernuzzi:2011aj}, we expect that the
resummed-factorized energy-flux and mode amplitudes agree quite well
with the Teukolsky data at least up to the ISCO, provided that the
spin is not too high. In fact, Ref. \cite{Pan:2010hz} showed that, in the
adiabatic limit, the resummed-factorized $(2,2)$ modes agree very well
with frequency-domain Teukolsky modes up to the ISCO, at least over
the range $-1 \le a/M \lesssim 0.7$.  The relative difference between
amplitudes in the two models is less than $0.5\%$ when $a/M\leq 0.5$,
but grows to $3.5\%$ when $a/M \simeq 0.75$.  

In this work, we focus
on the $(2,2)$ mode comparison, leaving to a future publication a
thorough study of the higher modes.  We have already seen in
Sec.~\ref{sec:characteristics} that as $a/M \rightarrow 1$, many more
modes become excited during the plunge and merger.  For this limit,
the resummed-factorized waveforms will need to be improved in order to
match higher-order Teukolsky modes with good precision.

\begin{table}
\centering 
\begin{tabular}{c| c| c| c| c| c}
\hline\hline 
$a/M$ & $\Delta t^{22}_{\rm peak}$ & $|h_{22,\rm peak}^{\rm Teuk}|$ & $d^2{|h_{22,\rm peak}^{\rm Teuk}|}/dt^2$ & 
$\omega_{22,\rm peak}^{\rm Teuk}$ & $\dot{\omega}_{22,\rm peak}^{\rm Teuk}$\T \B\\ 
\hline 
 -0.9 & 1.60 & 0.001341 & -3.532$\times10^{-6}$ & 0.2195& 0.005676  \\
 -0.5 & -0.08  & 0.001382 & -2.536$\times10^{-6}$ & 0.2376 &0.006112 \\
  0.5 & -7.22 & 0.001542& -1.334$\times10^{-6}$ & 0.3396 & 0.005095 \\
 0.7 & -12.77 & 0.001582 & -1.212$\times10^{-6}$ & 0.3883& 0.004068\\
0.9 & -39.09 & 0.001576 & -8.102$\times10^{-8}$ &0.4790 & 0.001779 \\
\hline\hline
\end{tabular}
\caption{\label{tab:summary_spin} Input values for the RHSs of
  Eqs.~(\ref{cond1})--(\ref{cond5}) for the EOB $(2,2)$ mode for
  spinning black-holes used in Figs.~\ref{h22-Cal-Comparisons-a05}
  ($a/M=\pm0.5$) and \ref{h22-Cal-Comparisons-a07} ($a/M=0.7$ and
  $a/M=-0.9$).  We also include data for the case $a/M= 0.9$, although
  we do not compare waveforms for this example.}
\end{table}

Figures \ref{h22-Cal-Comparisons-a05} and
\ref{h22-Cal-Comparisons-a07} compare $(2,2)$ modes for the EOB and
Teukolsky waves for spin values $a/M = \pm 0.5$, $a/M = 0.7$ and $a/M
= -0.9$.  We build the full EOB waveform following the prescription
described in Sec.\ {\ref{sec:results_improved}}, using the input
parameters shown in Table~\ref{tab:summary_spin}. For the cases
$a/M=0.5$ and $a/M=0.7$, we also use a {\it pseudo} QNM (pQNM) (in
addition to the {\it standard} QNMs) as suggested in
Refs.~\cite{Buonanno:2009qa,Pan:2011gk}. A possible physical
motivation of these pQNMs follows.  The peak of the orbital frequency
comes from orbits that are very close to the light-ring
position~\cite{Barausse:2009xi}, which in turn corresponds nearly to
the peak of the effective potential for gravitational perturbations
\cite{ChandraDetweiler1975:kip,1985PhRvD..31..290M,Hod:2009td,Dolan:2010wr,Cardoso:2008bp}.
Therefore, before the orbital frequency peaks, the gravitational-wave
emission is dominated by the source of the Teukolsky equations
(i.e. by the particle); afterwards, the emission is dominated by the
black-hole's QNMs.  In the standard EOB approach, the waveform is a
superposition of QNMs already after the peak of the numerical
amplitude $t_{\rm peak}^{\ell m}$.  However, we have seen that this
precedes the peak of the orbital frequency by a considerable time
interval: $-\Delta t_{\ell m}\approx 12$--$40 M$ for $a/M=0.5$ and
$a/M=0.7$ (see Table~\ref{tab:summary_spin}).  To account for the
effect of the particle emission before the peak of the orbital
frequency, we therefore introduce a pQNM having frequency
$\omega_{22}^{\rm pQNM}=2 \Omega_{\rm max}$
(cf.~Table~\ref{tab:delay_spinning}) and decay time $\tau_{22}^{\rm
  pQNM}= -\Delta t^{22}_{\rm peak}/2$.  We included this pQNM only for
$a/M=0.5$ and $a/M=0.7$. For smaller spins, since $\Delta t^{22}_{\rm
  peak}$ is small, the pQNM would be short lived and would not alter
our results significantly.

As in the non-spinning case, phase and amplitude agreement are
excellent until the ISCO. The phase differences remain small during
the plunge, until merger, and grow up to $\sim 0.1$ rad during the
ringdown. The amplitude difference grows to larger values, $\sim
20\mbox{--}30\%$ through merger and ringdown, because of the
limitations of our procedure to attach the QNMs in the EOB waveforms
(see Ref.\ {\cite{Pan:2011gk}}, Fig.\ 3 and associated discussion).
The phase difference during the merger-ringdown for the case $a/M =
0.7$ is larger, because for larger and larger spins the
resummed-factorized waveforms~\cite{Pan:2010hz} perform less and less
accurately around and beyond the ISCO.  In the case $a/M = -0.9$, the
disagreement between the EOB and Teukolsky $(2,2)$ becomes large and
oscillatory during ringdown.  This is a consequence of the fact that
the oscillatory frequency behavior discussed in
Sec.\ \ref{sec:characteristics} is particularly strong in this case,
but we are not including the associated $(2,-2)$ QNMs in our EOB
model.

\begin{table}
\centering 
\begin{tabular}{c| c| c| c}
\hline\hline 
$a/M$ & $\Delta t_\mathrm{peak}^{21}$ & $\Delta t_\mathrm{peak}^{33}$ & 
$\Delta t_\mathrm{peak}^{44}$ \T \B\\ 
\hline 
 -0.9 &  14.26 & 3.81 & 5.41  \\
 -0.5 &  12.63 & 2.71 & 4.36 \\
  0.5 &  3.87 & -1.99 & 0.28 \\
 0.7 &  0.10 & -5.02 & -2.16 \\
0.9 &  -34.48 & -17.57 & -11.80 \\
\hline\hline
\end{tabular}
\caption{\label{tab:HOMinputs} The time delay $\Delta
  t_\mathrm{peak}^{21}$, $\Delta t_\mathrm{peak}^{33}$ and $\Delta
  t_\mathrm{peak}^{44}$ defined in Eq.~\eqref{delta_new} for
  $a/M=-0.9,-0.5,0.5,0.7$ and $0.9$. Time delay information for the
  non-spinning case $a/M=0$ and the dominant $(2,2)$ mode are given in
  Table~\ref{tab:delay_nonspinning} and
  Table~\ref{tab:delay_spinning}.}
\end{table}

\begin{figure}
 \includegraphics[width=8cm,clip=true]{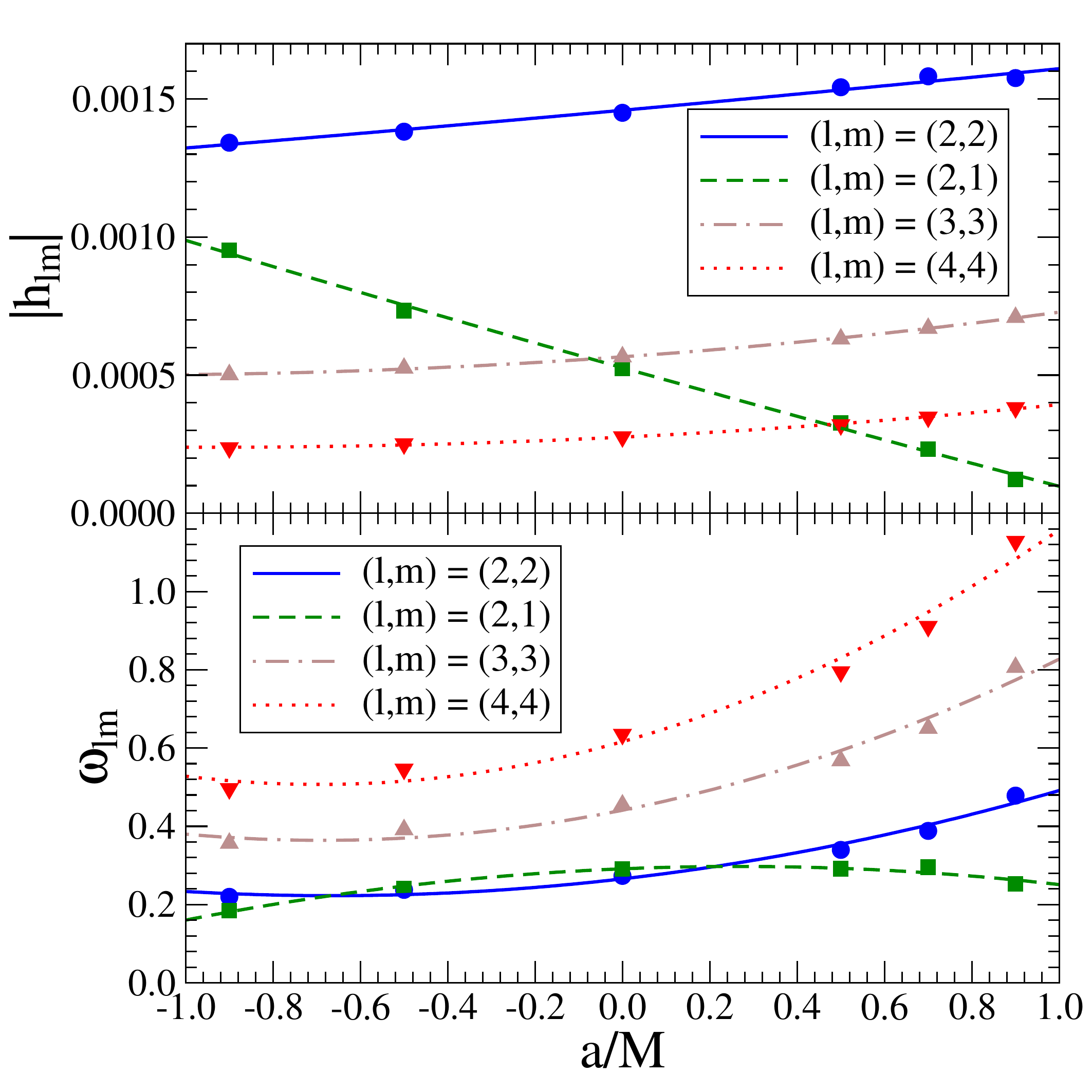} 
\caption{\label{fig:HOMinputs} Peak amplitude and corresponding
  frequency of Teukolsky modes as a function of spins.  Top panel is
  the peak amplitude $|h_{\ell m,\rm peak}^\mathrm{Teuk}|$; bottom is
  the frequency $\omega^\mathrm{Teuk}_{\ell m,\rm peak}$ extracted
  when the amplitude reaches this peak.  Data is for the $(2,2)$,
  $(2,1)$, $(3,3)$ and $(4,4)$ modes, for spins
  $a/M=-0.9,-0.5,0,0.5,0.7$ and $0.9$.  Solid lines are quadratic fits
  to these; the fits are given in Eq.~\eqref{HOMinputfits}.}
\end{figure}

Finally, although in this paper we do not attempt to calibrate
higher-order modes for $a \ne 0$, it is useful for ongoing work on the 
comparable-mass case to extract relevant information, such as the time
delay between the peaks of the $(\ell,m)$ modes, and the input
parameters entering the RHS of Eqs.~(\ref{cond1})--(\ref{cond5}).  In
Table \ref{tab:HOMinputs}, we show the time delays $\Delta
t_\mathrm{peak}^{\ell m}$; in Fig. \ref{fig:HOMinputs}, we show
$\left|h_{\ell m,\rm peak}^\mathrm{Teuk}\right|$ and $\omega^\mathrm{Teuk}_{\ell m,\rm peak}$
as functions of $a/M$.  Quadratic fits to these functions are
as follows:
\begin{eqnarray}\label{HOMinputfits}
\left|h_{22,\rm peak}^\mathrm{Teuk}\right| &=& 0.001\left[1.46 + 0.144\,a/M + 0.00704(a/M)^2\right] ,\nonumber\\
\left|h_{21,\rm peak}^\mathrm{Teuk}\right| &=& 0.001\left[0.527 - 0.445\,a/M + 0.016(a/M)^2\right] ,\nonumber\\
\left|h_{33,\rm peak}^\mathrm{Teuk}\right| &=& 0.001\left[0.566 + 0.133\,a/M + 0.0486(a/M)^2\right] ,\nonumber\\
\left|h_{44,\rm peak}^\mathrm{Teuk}\right| &=& 0.001\left[0.276 + 0.0773\,a/M + 0.0405(a/M)^2\right] ,\nonumber\\
\omega^\mathrm{Teuk}_{22,\rm peak} &=& 0.266 + 0.129\,a/M + 0.0968\,(a/M)^2 \,,\nonumber\\
\omega^\mathrm{Teuk}_{21,\rm peak} &=& 0.291 + 0.0454\,a/M - 0.0857\,(a/M)^2 \,,\nonumber\\
\omega^\mathrm{Teuk}_{33,\rm peak} &=& 0.441 + 0.224\,a/M + 0.163\,(a/M)^2 \,,\nonumber\\
\omega^\mathrm{Teuk}_{44,\rm peak} &=& 0.616 + 0.315\,a/M + 0.227\,(a/M)^2 \,.
\end{eqnarray}
We postpone to future work the study of $d^2\left|h_{\ell m,\rm
  peak}^\mathrm{Teuk}\right|/dt^2$ and
$\dot{\omega}^\mathrm{Teuk}_{\ell m,\rm peak}$ for higher-order modes.

\section{Conclusions}
\label{sec:concl}

The similarity of the transition from inspiral to merger to ringdown
over all mass ratios studied in Refs.~\cite{Buonanno00,Buonanno06}
suggested the possibility of using the test-particle limit as a
laboratory to investigate quickly and accurately the main features of
the merger signal.  The authors of
Refs.\ \cite{Nagar:2006xv,Damour2007} were the first to exploit this
possibility.  They proposed using the EOB inspiral-plunge trajectory
to build the source for the time-domain Regge-Wheeler-Zerilli
equations.  They also improved the EOB modeling, notably the energy
flux and the non-quasi-circular orbit effects, by requiring that the
EOB and RWZ leading $(2,2)$ mode agreed during plunge, merger and
ringdown.

\begin{figure}
 \includegraphics[width=8cm,clip=true]{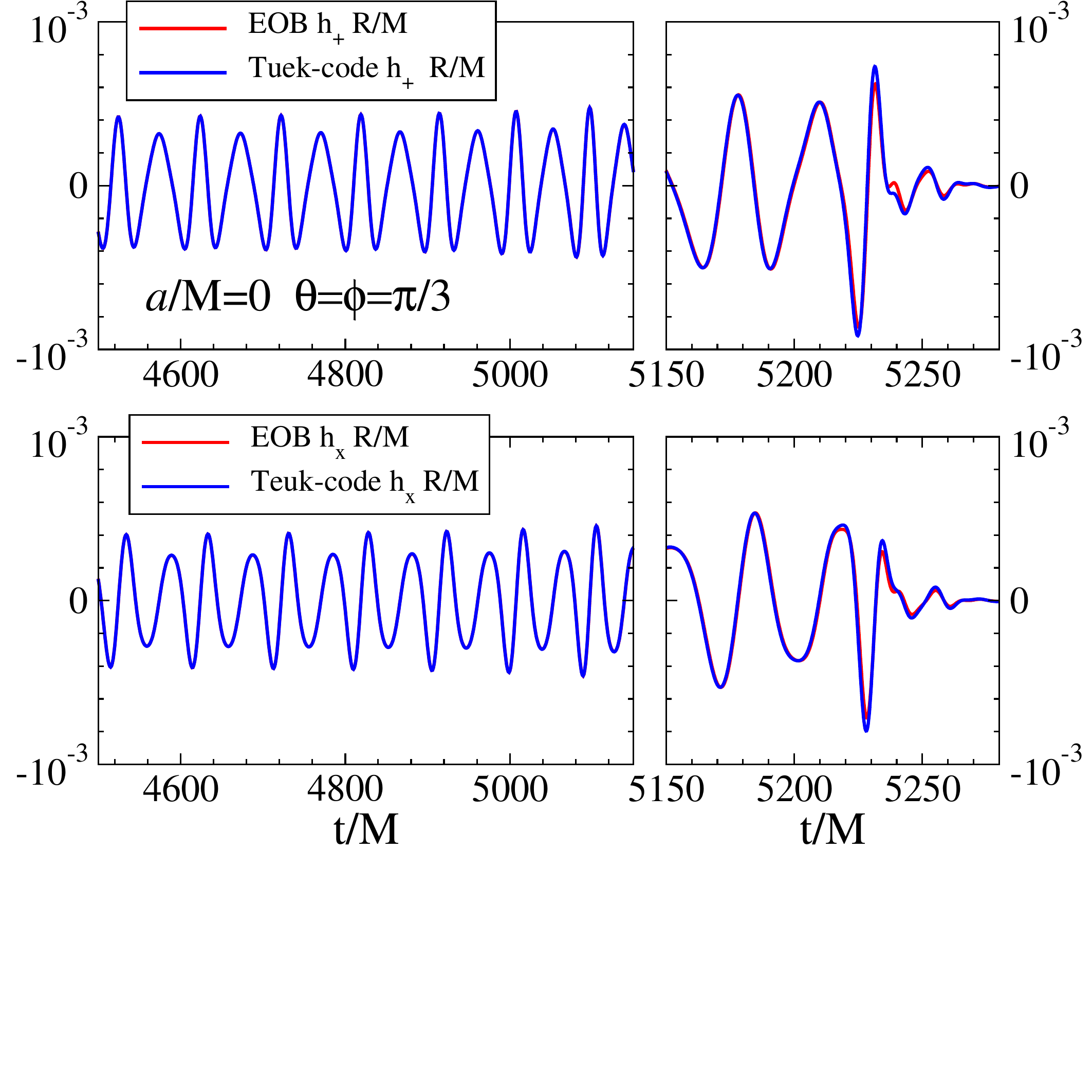} 
\vspace{-1.5cm}
\caption{\label{fig:hp-hc-00} Comparison between Teukolsky-calibrated
  EOB and Teukolsky $h_+$ and $h_\times$ polarizations for $a/M =0$.
  The four dominant modes $(2,2)$, $(2,1)$, $(3,3)$ and $(4,4)$ are
  included.}
\end{figure}

Here, we have employed the time-domain Teukolsky code developed in
Refs.~\cite{skh07,skh08, Sundararajan:2010sr} and extended previous
works~\cite{Nagar:2006xv,Damour2007,Bernuzzi:2010xj,Bernuzzi:2011aj}
in several directions.  In the Schwarzschild case, we first discussed
how the EOB model developed in Ref.~\cite{Pan:2011gk} for
comparable-mass non-spinning black holes performs when $\nu = 10^{-3}$
for the leading $(2,2)$ mode, as well as for three subleading modes,
$(2,1)$, $(3,3)$ and $(4,4)$. Confirming previous
results~\cite{Nagar:2006xv,Damour2007,Pan:2010hz,Fujita:2010xj,
  Bernuzzi:2010ty,Bernuzzi:2010xj,Fujita:2011zk,Bernuzzi:2011aj}, we
found that the agreement between the Teukolsky and EOB modes is
excellent during the long inspiral.  During the merger, whereas the
agreement of the $(2,2)$ and $(3,3)$ modes is still good, that of the
$(4,4)$ and $(2,1)$ is not very satisfactory.  We find that this is
due to the irregular behavior of the numerical-relativity input values
for the peak of the mode amplitude and the gravitational frequency at
that peak.  This motivates the need for more accurate
numerical-relativity data for these higher-order modes, which will
presumably be available in the future.  By calibrating the EOB model
using input values directly extracted from the Teukolsky modes
(Tables~\ref{tab:delay_nonspinning} and \ref{tab:delay_spinning}), we
found very good agreement for the four largest modes.  In
Fig.~\ref{fig:hp-hc-00}, we compare $h_+$ and $h_\times$ constructed
for these four modes, using
\begin{equation}\label{eq:Ylm-expansion}
h_+(\theta,\phi,t ) - i h_\times(\theta,\phi,t) = \sum_{\ell, m}
{}_{-\!2}Y_{\ell m}(\theta,\phi)\, h_{\ell m}(t)\,.
\end{equation}
The sum here is over $(\ell,m)=(2,\pm 2),(2,\pm 1),(3,\pm 3)$ and
$(4,\pm 4)$. The agreement between EOB and Teukolsky polarizations is
very good as expected. There are some minor differences during the
ringdown, which are mainly due to the underestimated ringdown
amplitudes of the $(2,2)$ and $(3,3)$ modes in the EOB model.

Moreover, for the first time, we employed the EOB inspiral-plunge
trajectory to produce merger waveforms for quasi-circular, equatorial
inspiral in the Kerr spacetime.  The energy flux in the EOB equations
of motion uses the factorized resummed waveforms of
Refs.~\cite{DIN,Pan:2010hz}.  We calibrated the leading EOB $(2,2)$
mode for spins $a/M = -0.9, -0.5, 0.5, 0.7$, and extracted information
on the subleading modes.  We also investigated the high spin case
$a/M=0.9$.  We found that several modes which are subleading during
the inspiral become relevant during plunge and merger.  The major new
feature of the EOB calibration (based on Teukolsky data) is that we
relaxed the assumption used in previous papers
\cite{Buonanno-Cook-Pretorius:2007, Buonanno2007, Pan2007,
  Buonanno:2009qa, Pan:2009wj, Damour2007a,
  DN2007b,DN2008,Damour2009a,Pan:2011gk} that the matching of the QNMs
for the leading $(2,2)$ mode occurs at the peak of the orbital
frequency.  In fact, we found that the peak of the orbital frequency
does not occur at the same time as the peak of the Teukolsky $(2,2)$
mode, and that the time difference grows as the spin parameter
increases.  Our work represents a first step in exploring and taking
advantage of test-particle limit results to build a better spin EOB
model in the comparable-mass case~\cite{Taracchini:2011}.

In the future, we plan to extend this work in at least two directions.
First, we want to calibrate the EOB model in the test-particle limit
for higher spins and for higher-order modes, and to connect it to the
spin EOB model in the comparable-mass case
\cite{Pan:2009wj,Taracchini:2011}.  To achieve this goal, we would
need to introduce adjustable parameters in the functions $\rho_{\ell
  m}$ in Eq.~(\ref{hlm}) to improve the resummed-factorized energy
flux and amplitude modes for large spin values.  

In our future
analyses, we will use a Teukolsky code which uses hyperboloidal slicing
\cite{Bernuzzi:2011aj,ZenKha2011}.  Although we were able to achieve
similar accuracy by extrapolating our results from finite radius to
future null infinity, hyperboloidal slicing is far faster, and has
proven to be very robust.  Second, we would like to extend this model
to inclined orbits.  To tackle this case, we need to generalize the
resummed-factorized waveforms to generic spin orientations.  If we
were only interested in extracting the input values, as in
Tables~\ref{tab:input}, \ref{tab:summary_spin}, it might be sufficient
to use the hybrid method suggested in Ref.~\cite{Han:2011qz}.  In this
case, we could use in the EOB equations of motion the energy flux
computed with a frequency-domain Teukolsky code \cite{h00}, but extend
it to plunging trajectories.

Finally, besides improving the EOB model, the possibility of
generating quickly and accurately merger waveforms in the
test-particle limit will allow us to investigate several interesting
phenomena, such as the distribution of kick velocities for spinning
black-hole mergers~\cite{Sundararajan:2010sr}, the energy and
angular-momentum released when a test particle plunges into a Kerr
black hole~\cite{kipOT,Kesden:2011ma}, and the {\it generic} ringdown
frequencies suggested in Refs.~\cite{MinoBrink2008,Zimmerman:2011dx}.

\section{Acknowledgements}

E.B., A.B. and Y.P. acknowledge support from NSF Grant PHY-0903631.
A.B. also acknowledges support from NASA grant NNX09AI81G.  GK
acknowledges research support from NSF Grant Nos. PHY-0902026,
CNS-0959382, PHY-1016906 and PHY-1135664, and AFOSR DURIP Grant
No. FA9550-10-1-0354.  SAH was supported by NSF Grant PHY-0449884 and
NASA Grant NNX08AL42G; SO was also supported by NASA Grant NNX08AL42G.
Most of the numerical simulations needed for this work were performed
on Georgia Tech's Keeneland supercomputer under project number
UT-NTNL0036.

\bibliography{References/References}
\end{document}